\documentclass[twocolumn, amsmath,amssymb, aps]{revtex4-1}

\usepackage{soul}
\usepackage{cancel}
\usepackage{listings}
\usepackage{relsize}
\usepackage{graphicx}
\usepackage{amsmath}
\usepackage{xcolor}
\usepackage{filecontents}
\usepackage{listings}
\DeclareFontFamily{U}{mathx}{}
\DeclareFontShape{U}{mathx}{m}{n}{<-> mathx10}{}
\DeclareSymbolFont{mathx}{U}{mathx}{m}{n}
\DeclareMathAccent{\widehat}{0}{mathx}{"70}
\DeclareMathAccent{\widecheck}{0}{mathx}{"71}

\renewcommand{\selectlanguage}[1]{}

\newcommand{\subnumber}[1]{\begin{flushleft}\bf{#1}\end{flushleft}}

\begin{document}

\title{Modeling of thin liquid films with arbitrary many layers
}

\author{Tilman Richter $^1$$^*$}
\author{Paolo Malgaretti $^1$$^\dagger$}
\author{Jens Harting $^1$$^2$}
\affiliation{
$^1$ Helmholtz Institute Erlangen-N\"urnberg for Renewable Energy,
Forschungszentrum J\"ulich, Cauerstr.~1, 91058 Erlangen, Germany\\
$^2$ Department of Chemical and Biological Engineering and Department of Physics, Friedrich-Alexander-Universität Erlangen-Nürnberg, Cauerstr.~1, 91058 Erlangen\\
$^*$t.richter@fz.juelich.de, $^\dagger$p.malgaretti@fz-juelich.de}


\begin{abstract}
We propose the generalization of the thin film equation (TFE) to arbitrarily many immiscible liquid layers. Then, we provide different pathways for deriving the hydrodynamic pressure within the individual layers, showing how to understand the equation as a Cahn-Hilliard-type conservation equation and providing an algorithm to derive the associated Onsager Matrix. Furthermore, we employ a numerical solver based on the multilayer shallow water-lattice Boltzmann method (LBM) for two and three liquid layers in pseudo two and three dimensions to gain insights into the dynamics of the system and to validate the model. We perform a linear stability analysis and assess droplet equilibrium shapes. Furthermore, we compare the dynamics of the proposed thin film equation to full Navier-Stokes simulations and show the possible equilibrium states of the multilayer liquid thin film system. 
\end{abstract}

\maketitle

\section{Introduction}

Thin liquid films play a crucial role in a wide range of natural and industrial processes, from coating technologies to biological systems and microfluidics. The wetting and dewetting behavior of these films determines important properties such as stability, spreading, and pattern formation, influencing applications in fields like ink-jet printing, microelectronics, and tissue engineering~\cite{oron_long-scale_1997, craster_dynamics_2009, de_gennes_capillarity_2004, lohse_fundamental_2022, gao_inkjet_2017}. In particular, understanding the dynamics of multi-layered liquid films is essential for advancing technologies where precise control over layer interactions and surface properties is required, especially in wet-on-wet coating and printing \cite{hack_printing_2018}.

Multilayer liquid systems are found in diverse applications, such as the design of functional coatings, where films with tailored properties must coexist in stable configurations \cite{bishop_development_2020, liu_multilayer_2022}, or in biological settings, where layered films can model structures like membranes and biofilms\cite{trinschek_continuous_2017, wallmeyer_collective_2018, kusumaatmaja_intracellular_2021}. In these contexts, phenomena like wetting, dewetting-induced patterning, and stability against rupture are of fundamental importance. Moreover, the ability to control multi-layer film behavior can lead to advancements in areas such as wearable electronics, printed photovoltaics, and biomedical engineering~\cite{guo_inkjet_2013, gao_inkjet_2017, sun_inkjet_2018, qiu_situ_2024, neto_novel_2007, bishop_development_2020}.

Early experimental studies, such as those by Hildebrand and Kittsley~\cite{hildebrand_seven_1949, kittsley_eight_1950}, demonstrated that liquid films can be composed of 8 or more immiscible layers, laying the groundwork for investigating more complex film configurations.

Despite the broad relevance of multi-layer films, much of the existing research has focused on the dynamics of single or two-layer systems. Subsequent theoretical work, particularly within the lubrication approximation framework, has offered insights into the behavior of thin liquid films and their stability~\cite{oron_long-scale_1997, craster_dynamics_2009}.

Theoretical models have advanced significantly with the development of the thin film equation (TFE) for two-layer systems, incorporating effects like van der Waals interactions and spinodal dewetting, which help to explain non-equilibrium dynamics and instabilities in such systems~\cite{danov_stability_1998, pototsky_alternative_2004, pototsky_evolution_2006, bommer_droplets_2013}. Recent work by Diekmann and Thiele~\cite{diekmann_mesoscopic_2024} has revisited these models, focusing on deriving the TFE through the variation of macroscopic and mesoscopic free energy formulations. These efforts have provided a deeper understanding of two-layer dynamics, but extending the analysis to systems with many layers remains a challenging and underexplored area.

In this work, we aim to extend the existing framework to multi-layer thin films. By deriving a comprehensive thin film equation for systems with arbitrary numbers of layers, we address the complexities that arise when multiple immiscible layers interact. Our approach is based on a first-principles derivation of hydrodynamic pressures, considering lubrication approximation, Stokes approximation, and the minimization of relevant free energies.

This paper unfolds as follows: Section~\ref{sec:Model} outlines the derivation of the evolution equations, emphasizing our novel approach; Section~\ref{sec:LBM} introduces a numerical solver based on the multilayer shallow water Lattice Boltzmann Method (LBM) for simulating the dynamics of multi-layer systems; Section~\ref{sec:Results} presents validation results and insights gained from numerical simulations of both two and three-layer systems. Further details and mathematical derivations are provided in the Appendix. 

This endeavor aims to provide a comprehensive framework for studying and understanding the dynamics of multi-layer thin liquid films, allowing for a simplified analytical and numerical treatment  due to the reduced dimensionality of the evolution equations. This will enable the future investigation of cutting-edge technologies concerned with wet-on-wet printing and coating.

\section{Model}
\label{sec:Model}

\begin{figure}
    \centering
    \includegraphics[width=\linewidth]{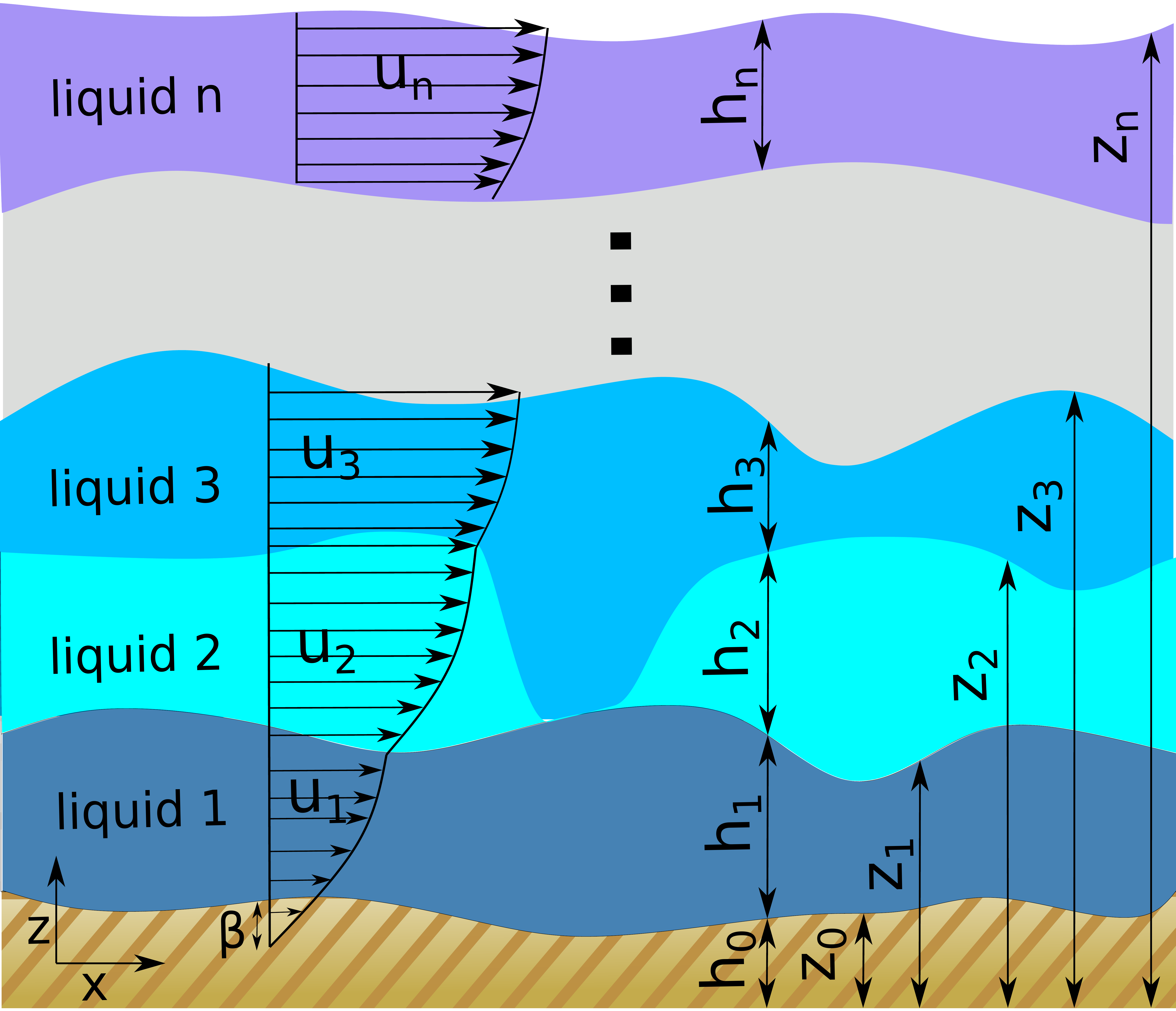}
    \caption{Sketch of the model. Shown are $n$-many layers of immiscible liquids, with their thicknesses $h_i$ and the interface position between the $i$th and the $(i+1)$th layer $z_i$. Index $0$ refers to the solid substrate. Further, we indicate a slip-length $\beta$ at the solid substrate that can be chosen $\beta=0$ to restore the no-slip condition. Finally, we sketch the horizontal component of the fluid velocity in the $i$th layer labeled $u_i$. 
    }
    \label{fig:sketch_n}
\end{figure}

We develop a set of evolution equations for a system as sketched in Fig.~\ref{fig:sketch_n}. That is, we consider $n$ layers of pairwise immiscible fluids resting on a solid substrate. The densities of the liquids are assumed in such a way that the densest liquid is liquid 1, and the density of the liquids reduces with their number, resulting in no two liquids changing their position in the stack. We will describe the different liquid layers by the vertical position of their interface $z_i$, where index $0$ indicates the possible corrugated solid substrate. Equivalently, one can use the thicknesses of the respective layers $h_i:=z_i-z_{i-1}$. Using both variables will help to keep the notation compact. We assume the position of the interfaces to vary only moderately in $\mathbf{x}$, i.e., $\nabla z_i\ll 1\forall i$, such that there is a length scale separation between horizontal and vertical length scales. This is called the lubrication approximation and is often employed in the  context of wetting~\cite{oron_long-scale_1997, craster_dynamics_2009}. Furthermore, we introduce the horizontal velocity $u=u_i(\mathbf{x},z)$ in the $i$th liquid layer as well as the vertical component of the velocity $v=v(\mathbf{x},z)$. We allow for a so-called precursor film of thickness $h^*$ for every liquid layer. The implication of this is that the thickness of every layer attains its minimum at a thickness of $h_i=h^*$. Even though there is evidence for the existence of a physical precursor film of molecular thickness in many experiments~\cite{popescu_precursor_2012} the precursor film in this model rather has to be understood as a necessity to avoid a divergence at the contact line of droplets. Finally, we introduce the possibility of slip with slip length $\beta$. The details of this will be described later on. For now, it is sufficient to state that choosing $\beta=0$ restores the classical no-slip condition while $\beta>0$ grants the liquid to slip over the layer/substrate below. 

\subsection{$n$ many thin layers of liquid}

\subsubsection{From the Navier-Stokes to the Thin-Sheet equation}

The starting point of our derivation is the compressible Navier-Stokes equations (NSE), consisting of mass and momentum conservation 
\begin{subequations}
\label{eq:NSE}
    \begin{align}
          \label{eq:NSE1}
    \nabla \cdot (u,v)&=0\\
    \label{eq:NSE2}
    \rho(\partial_t (u,v) + (u,v)\cdot \nabla (u,v))+\nabla p&= \mu \Delta (u,v).
    \end{align}
\end{subequations}
In Eq.~\eqref{eq:NSE}, we introduce the density of the liquids $\rho$ to be $\rho=\rho_i$ in the $i$th layer and the pressure $p=p_i$ in the respective layer. Following the lubrication approximation, we assume the pressure to be homogeneous along the transversal direction within a liquid layer $\nabla_z p_i =0$. Furthermore, we introduce the viscosity $\mu$ to be chosen $\mu=\mu_i$ in the respective layer. 

Using the Leibnitz rule $ \int_a^b \partial_x f dz = \partial_x \int_a^b f dz - \partial_x b f(b) + \partial_x a f(a)$ and the free boundary condtion $ \partial_t z_i(x,t)+ u(x,z,t)\nabla_x z_i(x,t)=v(x,z_i,t)$ we apply the integral $\int_{z_{i-1}}^{z_i}dz$ to the NSE Eq.~\eqref{eq:NSE}. 

The integration of Eq.~\eqref{eq:NSE1} yields 
\begin{align}
\label{eq:thin_sheet_1}
    \partial_t h_i + \nabla \cdot \left( h_i U_i\right) = 0.
\end{align}
In Eq.~\eqref{eq:thin_sheet_1} we define the flux $U_i:=\frac{1}{h_i}\int_{z_{i-1}}^{z_i} u dz$. 
The integration of Eq.~\eqref{eq:NSE2} leads to 
\begin{align}
    \label{eq:thin_sheet_2_prev}
     \rho_i( \partial_t(h_iU_i) + \nabla_x \eta_i h_i U_i^2 )=- h_i\nabla_x p_i+\mu_i\left. \partial_z u\right|_{z_{i-1}}^{z_i},
\end{align}
which includes the term $\eta_i := \frac{1}{h_i U_i^2}\int_{z_{i-1}}^{z_i} u^2 dz$. 
Applying the Stokes approximation $Re=\frac{\rho u L}{\mu}\ll 1$, the left hand side of Eq.~\eqref{eq:thin_sheet_2_prev} can be dropped and it becomes 
\begin{align}
    \label{eq:thin_sheet_2}
    0=- h_i\nabla_x p_i+\mu_i\left. \partial_z u\right|_{z_{i-1}}^{z_i}.
\end{align}
Hence, the factor $\eta$ is irrelevant in the viscous regime. From now on, we refer to Eqns.~$\eqref{eq:thin_sheet_1}$ and \eqref{eq:thin_sheet_2} as the \emph{Thin Sheet Equation}. Refs.~\cite{zitz_lattice_2019} and \cite{trontin_conservative_2020} use a single-layer version of those equations. In the ongoing discussion, we will omit the subscript $\mathbf{x}$, indicating derivatives in the horizontal directions only. 

The Thin Sheet Equation contains two terms that are so far unknown to us: the pressure $p$ and the friction force $\mu_i \left. \partial_z u \right|_{z_{i-1}}^{z_i}$, which are worked as follows.

\subsubsection{Boundary conditions and friction force}

Here, we derive the friction force $\mu_i \left. \partial_z u \right|_{z_{i-1}}^{z_i}$. We refer to it as a friction force as it stems from the dissipative term $\mu\nabla^2 u$ in the NSE and implements the slip/no-slip condition with the neighbouring liquid layers or the substrate. 

By the lubrication approximation, we assume that $p_i$ relaxes very fast and thus is assumed constant in the vertical direction and only varies in the horizontal direction. Therefore, we can conclude that the velocity $u$ in every liquid layer is given by a Poiseuille flow of quadratic shape $u=u_i$. With the ansatz 
\begin{align}
    \label{eq:u_ansatz}
    u_i(z)=a_iz^2+b_iz+c_i, 
\end{align}
the coefficients $a_i,b_i,$ and $c_i$ can be determined by formulating appropriate boundary conditions. 

We formulate the first boundary condition for the interface below the $i$th liquid layer. It is the continuity of the horizontal velocity where one can include slip at the interface if needed: 
\begin{align}
    \label{eq:no_slip}
    u_i(z_{i-1}-\beta_i)=u_{i-1}(z_{i-1}).
\end{align}
In Eq.~\eqref{eq:no_slip}, we imply $u_0(z_0)=0$, assuring a no-slip boundary condition at the solid substrate for $\beta_1=0$. 

The second condition that applies on $u_i$ is trivially the definition of the flux $U_i$,
\begin{align}
\label{eq:flux}
    \int_{z_{i-1}}^{z_i} dz =h_{i}U_i.
\end{align}
This, of course, introduces the flux $U_i$ into the thin sheet equation. We will, later on, see that this is very convenient as the flux will cancel out, leaving us with nothing but an evolution equation for the film thicknesses $h_i$ or equivalent, the interface positions $z_i$. 

Finally, the boundary condition for the upper interface of the $i$th liquid layer is given by the continuity of the stress tensor, which in the present context reads 
\begin{align}
\label{eq:viscous_boundary}
    \partial_z u_i(z_{i})=\frac{\mu_{i+1}}{\mu_i}\partial_z u_{i+1}(z_i). 
\end{align}
At the uppermost liquid layer, i.e., at the interface between liquid $n$ and the surrounding gas phase, we set $\mu_{n+1}=\mu_{atm}=0$.

Eqns.~\eqref{eq:no_slip}, \eqref{eq:flux}, and \eqref{eq:viscous_boundary} provide $3n$ linear independent equations, determining the $3n$ unknown parameters $a_i,b_i,$and $c_i$. Owing those parameters, we can derive that the friction force reads 
\begin{align}
    \mu_i \left. \partial_z u_i \right|_{z_{i-1}}^{z_i}= 2\mu_i h_i a_i.
\end{align}

All that is left to do to obtain the friction force is to solve the linear system 
\begin{align}
    \label{eq:boundary_conditions}
    D\begin{pmatrix} a_1\\ b_1 \\ c_2\\ \vdots \\ a_n \\ b_n\\ c_n\end{pmatrix} = \begin{pmatrix}
        0\\  z_1 U_1\\ 0 \\ \vdots\\ 0 \\ (z_{n-1}-z_n)U_n\\ 0 
    \end{pmatrix} 
\end{align}
for $a_i$, where the $3n\times 3n$ matrix $D$ is given by 
\begin{widetext}
    \begin{align}
     D := \left(
    \begin{array}{ccccccccccccc}
        \beta_1^2 & \beta_1 & 1 &  &  &  &  &  & & & & & \\
        \frac{z_1^3}{3} & \frac{z_1^2}{2} & z_1 &  &  & &  &  & & & & &\\
        2z_1 & 1 & 0 & -2\frac{\mu_2}{\mu_1}z_1 & - \frac{\mu_2}{\mu_1 } & 0 &  &  & & & & &\\
        z_1^2 & z_1 & 1 &  -(z_1-\beta_2)^2 & -(z_1-\beta_2) & -1 &  &  & & & & &\\
         0 & 0  & 0 & \frac{z_2^3-z_1^3}{3 }& \frac{z_2^2-z_1^2}{2} & z_2-z_1 &  &  & & & & &\\
          & & & & & & \ddots & & & & & &  \\
         &  &  & & & & & 2z_{n-1} & 1 & 0 & -2\frac{\mu_{n-1}}{\mu_n} z_{n-1} & -\frac{\mu_{n-1}}{\mu_{n}} & 0 \\
         &  &  & & & & & z_{n-1}^2 & z_{n-1} & 1 & -(z_{n-1}-\beta_n)^2 & - (z_{n-1}-\beta_n) & -1 \\
         &  &  & & & & & 0 & 0 & 0 & \frac{z_n^3-z_{n-1}^3}{3} & \frac{z_{n}^2 - z_{n-1}^3}{2} & z_n-z_{n-1}\\
         &  &  & & & & & 0 & 0 & 0 & 2z_n & 1 & 0 
    \end{array}\right).       
    \end{align}
\end{widetext}

Let $D^{-1}$ be the inverse of the matrix $D$. We then define the $n\times n$ matrix $S$ as the following entries of $D^{-1}$:
\begin{align}
     S=(D^{-1}_{3k+1, 3k+2})_{k=0, \cdots, n-1}.
\end{align}
With this, we calculate
\begin{align}
    \label{eq:friction_force}
    (2\mu_i h_i a_i)_{i=1,\cdots,n}= 2 \begin{pmatrix}
        \mu_1 h_1 & & \\
         & \ddots & \\
          & & \mu_n h_n 
    \end{pmatrix}S\begin{pmatrix}
        h_1 U_1 \\ \vdots \\ h_n U_n.
    \end{pmatrix}
\end{align}

Inverting the matrix $D$ provides some technical difficulties with the increasing number of considered liquid layers. Already for three layers, obtaining the friction force without the help of a computer algebra system is almost impossible. We want to stress that Eq.~\eqref{eq:boundary_conditions} is nothing but Eq.~\eqref{eq:no_slip}, Eq.~\eqref{eq:flux}, and Eq.~\eqref{eq:viscous_boundary} written into a compact matrix notation. Ref.~\cite{pototsky_morphology_2005} follows a similar line of arguments to obtain the Onsager matrix of a two-layer thin film system. 

\subsubsection{Pressure}

Next, we formulate the pressure $p_i$ in the $i$th liquid layer. We provide the pressure jump $p_{i}-p_{i+1}$ across the $i$th interface, which then trivially  allows to obtain the pressure by $p_i-p_{atm} = \sum_{j=i}^n p_{j}-p_{j+1}$. Here, $p_{atm}$ denotes the atmospheric pressure which is assumed  to be constant in time and homogeneous in space. 

There are two main contributions for the pressure jump to be accounted for, namely the Laplace pressure $\Lambda_i$ and the disjoining pressure $\Pi_i$, which leads to 
\begin{align}
    p_{i}-p_{i+1}=\Lambda_i + \Pi_i.
\end{align}
The Laplace pressure equals minus the local surface tension times the local curvature~\cite{butt_physics_2003,de_gennes_capillarity_2004, doi_soft_2013}. During the dynamics, some liquid layers may dewet, and hence, liquid layers that were previously far apart may come into contact. This implies that we must consider the dewetting dynamics when determining the surface tension a given liquid layer experiences. To do so, we write
\begin{align}
\label{eq:Laplace_pressure}
   \Lambda_i= -\left(\gamma_{i,i+1}+ \sum_{k=1}^{n-i}\varphi_{ik}(z_{i+k}-z_i)\right) \nabla^2 z_i. 
\end{align}
Eq.~\eqref{eq:Laplace_pressure} contains several new terms, which we introduce in the following. First, the surface tension between liquid $i$ and $i+k$ is denoted by $\gamma_{i,i+k}$. Next, we have added the interpolation function $\varphi_{ik}$, which interpolates between two limit cases. If all liquid layers between the interface $z_i$ and the interface $z_{i+k}$ are thick enough ($z_{i+1}-z_i\gg h^*$) there is no contact between, between liquid $i$ and liquid $i+k+1$. In this case, liquid $i$ is in contact with liquid $i+1$. If, on the other, all liquid layers between $z_i$ and $z_{i+k}$ (liquid $i+1$, liquid $i+2$, ..., liquid $i+k-1$, liquid $i+k$) dewet, then liquid $i$ and liquid $i+k+1$ come into contact. In Figure~\ref{fig:sketch_n}, the local dewetting of liquid 2 is sketched. Thus we need to replace the surface tension between liquid $i$ and liquid $i+1$ ($\gamma_{i,i+1}$) by the surface tension between liquid $i$ and liquid $i+k+1$ ($\gamma_{i,i+k+1}$). Also, all interaction for the in-between layers (layer $i$ with layer $i+2$, layer $i$ with layer $i+3$, ..., layer $i+1$ with layer $i+2$, ...) need to be canceled. Within our continuum model, we need to account for the fact that the effective surface tension will vary continuously while the film thickness is becoming vanishing small. We have chosen to use as the interpolating function the wetting potential\cite{de_gennes_capillarity_2004}, which indeed is a smooth function of the distance between the interfaces. Moreover, as shown in the Appendix, the wetting potential $\varphi$ naturally arises as part of the Laplace pressure when deriving the model by minimizing an effective free energy. Such derivations of thin film models from a free energy function are common in the thin film theory and have been employed to derive many thin-film models, including the standard TFE or the TFE including soluble surfactants \cite{thiele_gradient_2016} two-layer thin films \cite{pototsky_alternative_2004, pototsky_morphology_2005, pototsky_evolution_2006, diekmann_mesoscopic_2024}, or the wetting of elastic substrates \cite{henkel_gradient-dynamics_2021}.

A functional form that is often chosen throughout the literature is \cite{peschka_signatures_2019, henkel_gradient-dynamics_2021, diez_breakup_2007, zitz_lattice_2019, oron_long-scale_1997, craster_dynamics_2006, israelachvili_intermolecular_nodate} 
\begin{subequations}
\label{eq:wetting_potential}
\begin{align}
    \varphi_{ik}(h) &= S_{ik}\varphi_k(h)\\
   \varphi_k(h) &= \frac{1}{n-m}\left( n \left( \frac{kh^*}{h}\right)^m - m \left( \frac{kh^*}{h}\right)^n \right) 
\end{align}
\end{subequations}
where the index $ik$ indicates that the surface tension due to potential wetting of dewetting of $k$ layers starting at the $i$th interface is modeled. $n>m$ are two integers. The functional form of $\varphi_{ik}$ stems from integrating the pairwise Lennard-Jones potential between liquid molecules of distinct layers, giving an effective interface-interface potential. Employing the standard (6,12)-Lennard-Jones potential leads to $n=2$ and $m=8$ \cite{israelachvili_intermolecular_nodate, de_gennes_capillarity_2004, goldsteinsoft, peschka_signatures_2019}. In Eq.~\eqref{eq:wetting_potential} $h^*$ is the precursor height, i.e., the minimum film thickness allowed within our continuum model. Such a choice ensures the stability of the numerical solver. Accordingly, a film of thickness $h^*$ is considered dewetted. As $\varphi_{ik}$ models the potential dewetting of $k$ many layers, the factor $k$ in front of $h^*$ is necessary, as $kh^*$ is the thickness of $k$ many wetted layers on top of each other.  The physics of the problem does not crucially depend on the specific choice for $\varphi_{ik}$, and other choices are valid \cite{oron_long-scale_1997, craster_dynamics_2006}. 

Further, Eq.~\eqref{eq:wetting_potential} introduces the spreading 
coefficients
\begin{subequations}
\label{eq:spreading_parameters}
    \begin{align}
    S_{i1}=&\gamma_{i,i+2}-\gamma_{i,i+1}-\gamma_{i+1,i+2}\\
    S_{ik}=&\gamma_{i,i+k+1}+\gamma_{i+1,i+k}-\gamma_{i,i+k}-\gamma_{i+1,i+k+1}.
\end{align}
\end{subequations}
These are a carefully chosen sum of surface tensions cancelling out the right terms when the respective liquid layers dewet. When a number of liquid layers dewet, the associated liquid interfaces disappear, while new interfaces between liquids that were formerly separated by other liquid layers emerge. By summation over $i$ and $k$, Eq.~\eqref{eq:spreading_parameters} ensures to account for the correct surface tensions in all situations. We remark that $S_{i1}$ and $S_{ik}$ include surface tensions not indexed $i$. This is necessary because adjacent layers experience pressure jumps  $\Lambda_i, \Lambda_{i+1}, \cdots, \Lambda_{i+k}$, which need to cancel out upon dewetting. of the layers $i+1, \cdots, i+k$.

The disjoining pressure $\Pi_i$ is the attraction of two distant layers separated by yet different layers of liquid. It stems from pairwise interactions between the molecules of the attracted layers and decays very rapidly. The disjoining pressure can be observed for film thicknesses of at most several tens of nanometers~\cite{vrij_possible_1966, fetzer_thermal_2007}. Thus, the disjoining pressure is most important for liquid layers separated by only a small third layer and not so much across multiple layers. Nevertheless, we formally include the disjoining pressure between every possible combination of two liquid layers, including the substrate and the surrounding gas atmosphere, regardless of how small those contributions may be. They will be relevant once certain layers of liquid dewet. We must be careful to include both contributions of every pairwise attraction. The upper one of the layers will be pulled down, thus experiencing a positive pressure, while the lower layer will be pulled up, experiencing a negative pressure. 

The disjoining pressure is also the term that enforces the contact angle in the thin film equation. With possibly deformed interfaces, we have to account for the fact that the contact angle has to be taken with respect to the slope of the underlying interface.

This being said, we can formulate the disjoining pressure as 
\begin{align}
\label{eq:disjoining_pressure}
    \begin{split}
        \Pi_i = & -\sum_{k=1}^{n-i} \varphi_{ik}'(z_{i+k}-z_i)\left(1  +\frac{(\nabla z_i)^2}{2} \right)\\
        & + \sum_{l=0}^{i-1}\varphi_{l,i-l}'(z_i-z_l)\left(1 +\frac{(\nabla z_l)^2}{2} \right). 
    \end{split}
\end{align}
where $\sqrt{1+|\nabla z_i|^2}\approx 1+\frac{|\nabla z_i|}{2}$ is the surface element of the $i$th interface. The terms proportional to $(\nabla z_i)^2$ and $(\nabla z_l)^2$ ensure that the contact angle is enforced with respect to the local slope of the interface below. In the standard single-layer TFE (compare with Refs.~\cite{oron_long-scale_1997, craster_dynamics_2006}), that term is not present as the substrate is assumed to be flat ($z_0\equiv 0\Rightarrow \nabla z_0=0$). Further, in Eq.~\eqref{eq:disjoining_pressure}. $\varphi'_{ik}$ denotes the derivative of the wetting potential Eq.~\eqref{eq:wetting_potential}. $\phi'_{ik}$ is monotone decreasing around $kh^*$, and $\phi'_{ik}(kh^*)=0$. This ensures the desired minimum film height of $h^*$ for dewetted layers. 

The pressure $p_i$ in the $i$th liquid layer is given by $p_i-p_{atm}=p_i-p_{n+1}=\sum_{j=i}^n \Lambda_j+ \Pi_j$.

The pressure jump stated here can be derived as the variation of a free energy functional
\begin{align}
    p_i-p_{i+1}=\frac{\delta\mathcal{F}[h_1, \cdots, h_n]}{\delta h_i}.
\end{align}
The details of this derivation are given in the Appendix, where we also discuss the role of the wetting potentials $\varphi_{ik}$ in more detail.

\subsubsection{Thin film equation}

We have derived all necessary contributions and are ready to formulate the final evolution equation for the thickness $h_i$ of the $i$th layer. 
The thin sheet equation Eqns.~\eqref{eq:thin_sheet_1}, \eqref{eq:thin_sheet_2} reads
\begin{subequations}
\begin{align}
\label{eq:thin_sheet_matrix_1}
    \partial_t (h_1, \cdots , h_n)^T + \nabla\cdot (h_1 U_1 , \cdots, h_n U_n)^T=0
\end{align}
    \begin{align}
    \label{eq:thin_sheet_matrix_2}
        \begin{split}
             0=&-\begin{pmatrix}
            h_1 &\cdots & h_1 \\
             & \ddots &\vdots \\
              & & h_n
        \end{pmatrix} \nabla \begin{pmatrix}
            p_1-p_2 \\ \vdots \\ p_n-p_{atm}
        \end{pmatrix}\\ &+ 2 \begin{pmatrix}
        \mu_1 h_1 & & \\
         & \ddots & \\
          & & \mu_n h_n 
    \end{pmatrix}S\begin{pmatrix}
        h_1 U_1 \\ \vdots \\ h_n U_n
    \end{pmatrix}. 
        \end{split}
    \end{align}
\end{subequations}
We solve Eq.~\eqref{eq:thin_sheet_matrix_2} for $(h_1U_1, \cdots, h_nU_n)^T$ and insert it into Eq.~\eqref{eq:thin_sheet_matrix_1} yielding
\begin{align}
\label{eq:thin_film_equation_n}
    \partial_t\begin{pmatrix}
        h_1 \\ \vdots \\ h_n
    \end{pmatrix} =- \nabla\cdot \left( M \nabla \begin{pmatrix}
            p_1 - p_2\\ \vdots \\ p_n-p_{atm}
    \end{pmatrix}\right),
\end{align}
where the mobility matrix $M$ is given by
\begin{align}
\label{eq:mobility_n}
    M:=\frac{1}{2}\begin{pmatrix}
        \frac{1}{\mu_1 h_1} & & \\
        & \ddots & \\
        & & \frac{1}{\mu_n h_n}
    \end{pmatrix}S^{-1} \begin{pmatrix}
            h_1 &\cdots & h_1 \\
             & \ddots &\vdots \\
              & & h_n
        \end{pmatrix}. 
\end{align}
Eq.~\eqref{eq:thin_film_equation_n} is the general thin film equation for $n$ liquid layers. Explicitly writing it down requires inverting two matrices. First, the $3n\times 3n$ matrix $D$ needs to be inverted to obtain the matrix $S$. Then, the matrix $S$ must also be inverted. This process is complex and requires the use of a computer algebra system for three or more liquid layers.

\subsection{Two liquid layers}

In this section, we  explicitly state the thin film equation describing two immiscible layers of liquid resting on a homogenous solid substrate ($z_0=0$). It reads 
\begin{align}
\label{eq:thin_film_equation_2}
\partial_t \begin{pmatrix}
    h_1\\ h_2 
\end{pmatrix}  = \nabla\left( M \begin{pmatrix}
    \nabla(p_1-p_2) \\ \nabla (p_2-p_{atm})
\end{pmatrix} \right)  ,
\end{align}
with $p_{atm}$ being the atmospheric pressure assumed to be constant in time and homogeneous in space. The mobility matrix $M$ can be calculated from Eq.~\eqref{eq:mobility_n} and reads upon choosing a no slip boundary condition $\beta_1=\beta_2=0$ as
\begin{align}
\label{eq:mobility_2}
    M= \begin{pmatrix}
        \frac{z_1^3}{3\mu_1} & \frac{z_1^3 }{3\mu_1} +\frac{z_1^2 (z_2-z_1)}{2\mu_1} \\ 
       \frac{z_1^2(z_2-z_1)}{\mu_1} & \frac{(z_2-z_1) z_1^2 }{2\mu_1 }- \frac{(z_2-z_1)^2 (\mu_1 z_1 - \mu_1 z_2 - 3z_1 \mu_2)}{3\mu_1 \mu_2}
    \end{pmatrix}.
\end{align}
Rightfully, one may object that this matrix is not symmetric as the Onsager relation requires. However, this can be solved by  
a change of variables $(h_1,h_2)\mapsto(z_1,z_2)$ which leads to
\begin{align}
\label{eq:thin_film_pototski}
\partial_t \begin{pmatrix}
    z_1\\ z_2 
\end{pmatrix}  = \nabla\left( \tilde M \begin{pmatrix}
    \nabla(p_1-p_2) \\ \nabla (p_2-p_{atm})
\end{pmatrix} \right)  ,
\end{align}
where 
\begin{align}
\label{eq:mobility_pototski}
    \begin{split}
        \tilde{M}=& \begin{pmatrix}
        1 & 0 \\ 1 & 1
    \end{pmatrix} M\\
    =& \begin{pmatrix}
        \frac{z_1^3}{3\mu_1} & \frac{z_1^3}{3\mu_1}+ \frac{z_1^2(z_2-z_1)}{2\mu_1}\\
         \frac{z_1^3}{3\mu_1}+ \frac{z_1^2(z_2-z_1)}{2\mu_1} &\frac{(z_2-z_1)^3(\mu_1-\mu_2)}{3\mu_1\mu_2}+\frac{z_2^3}{3\mu_1}
    \end{pmatrix}.
    \end{split} 
\end{align}
Hence, the symmetry properties of the Onsager matrix are retrieved.
Eq.~\eqref{eq:thin_film_pototski} together with the mobility matrix Eq.~\eqref{eq:mobility_pototski} is the thin film equation formulated by Pototsky et al.~\cite{pototsky_alternative_2004, pototsky_morphology_2005, pototsky_evolution_2006} for two liquid layers showing that in the case $n=2$ we reproduce the model already known in the literature.  

We still have to provide the pressure explicitly for two liquid layers. It is given by 
\begin{subequations}
    \begin{align}
        \begin{split}
            p_1-p_2 =&  \varphi_{01}'(z_1)-\varphi_{11}'(z_2-z_1)\left(1+\frac{|\nabla z_1|^2}{2}\right)\\
            &-  (\gamma_{12}+\varphi_{11}(z_2-z_1)) \nabla^2 z_1
        \end{split}
    \end{align}
    \begin{align}
        \begin{split}
            p_2 - p_{atm} =& \varphi_{11}'(z_2-z_1)\left(1+\frac{|\nabla z_1|^2}{2}\right)\\
            &+\varphi'_{02}(z_2)  - \nabla\cdot \gamma_{2v} \nabla z_2
        \end{split}
    \end{align}
\end{subequations}
leading to the complete evolution equations for a two-layer thin film system. 

In the Appendix, we discuss obtaining the single-layer thin film equation as limit cases of the two-layer thin film equation Eq.~\eqref{eq:thin_film_equation_2}. 

\subsection{Three liquid layers}

For three liquid layers on a homogeneous solid substrate ($z_0=0$) the evolution equation Eq.~\eqref{eq:thin_film_equation_n} becomes 
\begin{align}
\label{eq:thin_film_equation_3}
\partial_t \begin{pmatrix}
    h_1\\ h_2 \\ h_3
\end{pmatrix}  = \nabla\left( M \nabla \begin{pmatrix}
    p_1-p_2 \\ p_2-p_{3} \\ p_3-p_{atm}
\end{pmatrix} \right)  ,
\end{align}
\\
\noindent where the mobility matrix $M$ is obtained from Eq.~\eqref{eq:mobility_n}. We do not write it down here explicitly as it is a rather long and cumbersome expression. It can be straightforwardly obtained through a computer algebra system (a code script is included in the SI). 

The pressure is given by 
\begin{widetext}
   \begin{subequations}
        \begin{align}
   \begin{split}
        p_1-p_2=&- \left(\gamma_{12}+\varphi_{11}(h_2)+\varphi_{12}(h_2+h_3)\right) \nabla^2 z_1 - \left((\varphi_{11}'(h_2)+\varphi_{12}'(h_2+h_3)\right)\left(1+\frac{(\nabla z_1)^2}{2} \right)\\
        & + \varphi'_{01}(h_1)\left(1  + \frac{(\nabla z_0)^2}{2}\right)
   \end{split}\\
   \begin{split}
       p_2-p_3=&-(\gamma_{23}+\varphi_{21}(h_3)) \nabla^2 z_2- \varphi'_{21}(h_3)\left(1+ \frac{(\nabla z_2)^2}{2}\right)+ \varphi'_{02}(z_2)\left(1+ \frac{(\nabla z_0)^2}{2}\right)\\
    &+\varphi'_{11}(h_2)\left(1+\frac{(\nabla z_1)^2}{2}\right)
   \end{split}\\
    \begin{split}
         p_3-p_{atm}=&-\gamma_{3v}\nabla^2 z_3 + \varphi'_{03}(z_3)\left(1 + \frac{(\nabla z_0)^2}{2}\right)  + \varphi'_{12}(h_2+h_3)\left(1  + \frac{(\nabla z_1)^2}{2}\right)\\
    &+ \varphi'_{21}(h_3)\left(1 + \frac{(\nabla z_2)^2}{2}\right).
    \end{split}
    \end{align}
   \end{subequations}
\end{widetext}

\section{Lattice Boltzmann scheme for multi-layer liquid films}
\label{sec:LBM}

\begin{figure}
    \subnumber{(a)}
    \includegraphics[width=\linewidth]{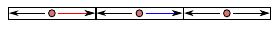}\\
    \subnumber{(b)}
    \includegraphics[width=\linewidth]{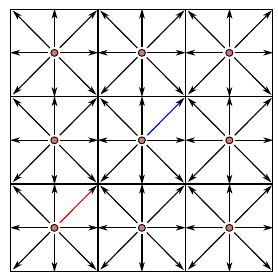}
    \caption{LBM lattices, showing the discretization of the phase space. We show discrete points in space (dots) with discrete directions of the momentum (arrows) attached to them. An example streaming step is indicated by the streaming of the according distribution function from the red to the blue arrow. (a) The D1Q3 Lattice for simulation of one-dimensional fields. (b) The D2Q9 Lattice for simulation of two-dimensional fields.}
    \label{fig:LBM_lattices}
\end{figure}

This section presents a modified lattice Boltzmann method (LBM) based on the shallow water equation to solve the evolution equation Eq.~\eqref{eq:thin_film_equation_n}. For a general introduction to the LBM, see Ref.~\cite{kruger_lattice_2017} or Ref.~\cite{succi2001lattice}. The original LBM was developed by discretising the Boltzmann equation of kinetic theory. By a Chapman-Enskog expansion, it can be shown that such an approach then naturally recovers the NSE. Despite classical fluid mechanics still being the most common use case of the LBM, it has been expanded in the last 30 years to solve many equations beyond the NSE. Examples of LBMs known to literature include integration schemes for Einstein's equations of general relativity~\cite{ilseven_lattice_2016}, the Maxwell equations of electrodynamics~\cite{mendoza_three_2010}, as well as for the shallow water equation~\cite{salmon_lattice_1999}, quantum mechanics~\cite{palpacelli_quantum_2008}, or wave-equation~\cite{chopard_lattice_1997, velasco_lattice_2019}. In the context of this work, the LBM for shallow water is especially interesting as it is the foundation of the method presented here. 

Many authors directly solve the thin film equation when modelling thin liquid films~\cite{pototsky_morphology_2005, lam_computing_2019}. This often requires large resolution or remeshing at the three-phase contact line of droplets~\cite{peschka_signatures_2019, henkel_gradient-dynamics_2021}. Further, such methods are, in general, difficult to implement and to parallelise. 


We follow a different route based on solving the shallow water equation and applying an appropriate forcing that ensures the needed dynamics. The numerical method presented here extends the method developed by Refs.~\cite{zitz_lattice_2019, zitz_swalbejl_2022}, modifying the shallow water equation for single-layer thin film dynamics. Ref.~\cite{trontin_conservative_2020} follows a similar approach employing a finite volume numerical scheme instead of an LBM using the same core idea. 

The central elements of an LBM are the discretized probability density function $f_{ki}$ in the momentum phase space. The phase space is discretized by considering a regular quadratic grid of points in space  with a number of velocity vectors, including the zero vector, attached to every single point. Here, we consider the two possible grids sketched in Fig.~\ref{fig:LBM_lattices}. For pseudo-two-dimensional simulations, we use a so-called D1Q3 lattice consisting of a one-dimensional array of evenly spaced points with the vectors
\begin{align}
    c_0=0, \quad c_1=1, \quad c_2=-1
\end{align}
attached to them~\cite{Qian1992}. On this lattice, we define an array of $3\cdot n $ discretized probability density functions $f_{ki}$  for every lattice point where the index $k$ corresponds to the three directions $c_k$ and the index $i$ to the $n$ liquid layers. For pseudo-three-dimensional simulations, we use the D2Q9 lattice, which consists of a two-dimensional array of evenly spaced points with the 9 unit vectors 
\begin{align}
\begin{split}
    &c_0=(0,0), \quad  c_1=(1,0), \quad c_2=\sqrt{2}(1,1)\\
    &c_3=(0,1),\quad   c_4=\sqrt{2}(-1,1), \quad  c_5=(-1,0)\\
    &c_6=\sqrt{2}(-1,-1), \quad c_7=(0,-1),  \quad c_8=\sqrt{2}(1,-1)
\end{split}
\end{align}
attached to them. On the D2Q9 lattice, we consider the $9\cdot n$ discretized probability density function $f_{ki}$  for every lattice point. 

The core of the algorithm is updating the density functions $f_{ki}$ by the Bhatnagar–Gross–Krook (BGK) collision and streaming operator
\begin{align}
\label{eq:BGK}
    \begin{split}
        &f_{ki}(x+c_k, t+\Delta t)\\
        &=f_{ki}(x,t)-\frac{1}{\tau}\left( f_{ki}(x,t)-f_{ki}^{eq}(h_i,U_i)\right)+ w_k \frac{1}{c_2^2}c_kF_i. 
    \end{split}
\end{align}
In Eq.~\eqref{eq:BGK}, we have the relaxation time $\tau$, i.e., the time in units of $\Delta t$ it takes the probability density function $f_{ki}$ to relax towards its local equilibrium $f^{eq}_{ki}$. For all practical purposes in this paper, one can choose $\tau=1$. The local equilibrium is given by 
\begin{subequations}
\label{eq:equilibrium_distribution}
    \begin{align}
        f_{k0}^{eq}=&h_i\left(1-w_0\frac{1}{2c_s^2}U_i^2\right) \\
        f_{ki}^{eq}=&w_k h_i \left(\frac{c_k\cdot U_i}{c_s^2}+\frac{(c_k\cdot U_i)^2}{2c_2^4}-\frac{U_i^2}{2c_s^2} \right). 
\end{align}
\end{subequations}
Furthermore, Eq.~\eqref{eq:BGK} and Eq.~\eqref{eq:equilibrium_distribution} contain the weights 
\begin{align}
    w_0=\frac{2}{3}, \quad w_{1,2}=\frac{1}{6}
\end{align}
for the D1Q3 lattice and 
\begin{align}
    w_0=\frac{4}{9}, \quad w_{1,3,5,7}=\frac{1}{9}, \quad w_{2,4,6,8}=\frac{1}{36}
\end{align}
the D2Q9 lattice. $c_s$ is the lattice speed of sound given by $c_s^2=1/3$. Finally, we specify the force $F_i$ to be given by 
\begin{align}
    \label{eq:force}
    F_i=-h_i\nabla p_i + 2\mu_i h_i a_i
\end{align}
where $2\mu_i h_i a_i$ is given by Eq.~\eqref{eq:friction_force}. Remark that Eq.~\eqref{eq:equilibrium_distribution} lacks the dispersive term of the standard LBM method \cite{kruger_lattice_2017}. Dissipation is introduced into the system by the friction force $2\mu_i h_i a_i$. 

Let us discuss in detail what Eq.~\eqref{eq:BGK} does. It consists of three parts. The collision term $f_{ki}^{col}=f_{ki}(x,t)-\frac{1}{\tau}\left( f_{ki}(x,t)-f_{ki}^{eq}(h_i,U_i)\right)$ where the probability density function $f_{ki}$ is relaxed towards the equilibrium. Then there is the forcing term $f_{ki}^{force}= w_k\frac{1}{c_s^2}F_i$. Finally, we stream the two terms discussed before into the respective direction $c_k$ by $f_{ki}(x+c_k, t+\Delta t)=f_{ki}^{col}+f_{ki}^{force}$. For example, the probability density functions associated with the red arrows in Fig.~\ref{fig:LBM_lattices} stream to the blue ones. 

From the probability density functions $f_{ki}$ we then obtain the thickness of the $i$th layer $h_i$ and the associated flux $U_i$ by summing 
    \begin{align}
        h_i=\sum h_{ki}, \quad h_iU_i \sum c_k f_{ki}.
    \end{align}

By a Chapman-Enskog analysis, one can show that $h_i$ and $U_i$ solve the shallow water equation to second order~\cite{salmon_lattice_1999},
\begin{subequations}
    \begin{align}
    \label{eq:shallow_water_1}
        \partial_t h_i + \nabla \cdot (h_i U_i)=&0\\
        \label{eq:shallow_water_2}
        \partial_t (h_iU_i) +\nabla  ( h_i U_i^2)=&F_i. 
    \end{align}
\end{subequations}
While Eq.~\eqref{eq:shallow_water_1} coincides with Eq.~\eqref{eq:thin_sheet_1}, Eq.~\eqref{eq:shallow_water_2} is only the same as Eq.~\eqref{eq:thin_sheet_2} when $\eta\approx1$, which is not necessarily the case. In fact, $\eta\approx 1$ only holds for a plug-flow ($\partial_z u =0$). This is contradictory to our model as we assume no slip or partial slip at the solid substrate (compare Fig.~\ref{fig:sketch_n}). As we have seen in section~\ref{sec:Model} and is also shown by Refs.~\cite{zitz_lattice_2019,trontin_conservative_2020}, the left-hand side of Eq.~\eqref{eq:shallow_water_2} can be neglected in the Stokes limit, rendering the disagreement between Eq.~\eqref{eq:shallow_water_2} and Eq.~\eqref{eq:thin_sheet_2} unimportant. Thus, the scheme described here solves the multi-layer-thin-film-equation Eq.~\eqref{eq:thin_film_equation_n} in the Stokes limit. As the multi-layer-thin-film-equation was derived using the Stokes limit and is only valid there, this is an acceptable shortcoming of the numerical method. 

To summarize this section, we have described an LBM scheme that solves multi-layer-thin-film-equation Eq.~\eqref{eq:thin_film_equation_n}. 

\section{Results}
\label{sec:Results}

This section presents numerical simulations from the method presented in the two preceding sections. We do so to verify our method and gain some insights into the equilibrium and dynamics of the system. We mainly focus on two liquid layers and two-dimensional simulations but also showcase three layers and three-dimensional simulations. 

\subsection{Linear stability analysis}
\label{sec:lsa}

\begin{figure}
\centering
\includegraphics[width=\linewidth]{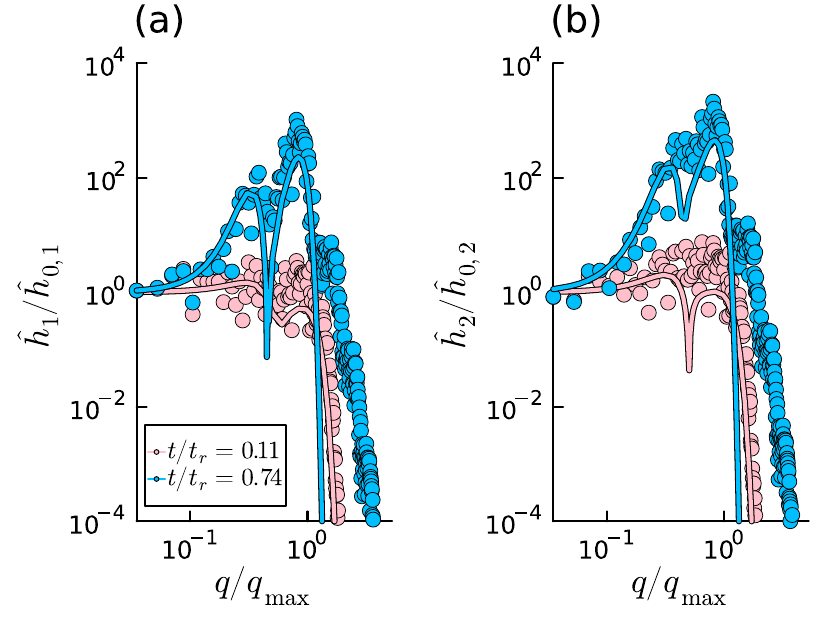}
\caption{Comparison of the linear stability analysis (LSA) with simulation data. Inset (a) shows the Fourier transform of the thickness of layer one $\widehat{h_1}$ together with the result predicted by the linear solution Eq.~\eqref{eq:linear_solution}. Inset (b) depicts the same for the Fourier transform of the upper layer thickness $\widehat{h_2}$. The colour code in both plots distinguishes the two selected time steps shown. Time is normalized by the rupture time of the system as measured from the simulation.
}
    \label{fig:lsa}
\end{figure}

As a first test case, we perform a linear stability analysis of the two-layer TFE (Eq.~\eqref{eq:thin_film_equation_2}) and compare it to the spectrum obtained by a fast Fourier transformation (FFT) of a time-evolution-simulation of a two-layer system as described in Sec.~\ref{sec:LBM}. 

We perform the linear stability analysis about the homogeneous state  $h_{0,1}$, $h_{0,2}$ to which we add a small perturbations
\begin{align}
\label{eq:lsa_ansatz}
    h_1=h_{0,1}+\delta h_1, \quad h_2=h_{0,2}+\delta h_2,
\end{align}
i.e., we assume that the thicknesses of both layers are perturbed by $\delta h_i$ around a constant value $h_{0,i}$. As a next step, we plug the ansatz Eq.~\eqref{eq:lsa_ansatz} into Eq.~\eqref{eq:thin_film_equation_2} and drop all terms that are of second order or higher in $\delta h_i$. When Fourier-transforming the obtained equation (where we denote the Fourier transform by a hat over the respective symbol), we obtain 
\begin{align}
\label{eq:linear_thin_film_equation}
    \partial_t \begin{pmatrix}
        \widehat{\delta h_1} \\ \widehat{\delta h_2} 
    \end{pmatrix}=-q^2M(h_{0,1},h_{0,2})E \cdot\begin{pmatrix}
        \widehat{\delta h_1} \\ \widehat{ \delta h_2}
    \end{pmatrix}.
\end{align}
In Eq.~\eqref{eq:linear_thin_film_equation} $q$ is the wave-number, stemming from the Fourier-transform, $M$ is the mobility matrix defined by Eq.~\eqref{eq:mobility_2} and the matrix $E=(E_i^j)_{i,j=1,2}$ is defined by 
\begin{align}
\begin{split}
    E_1^1=& \varphi_{01}''(h_{0,1})  + (\gamma_{12}+S_2\Theta_1(h_{0,2}))q^2\\
    E_1^2=& -\varphi_{11}''(h_{0,2})\\
    E_2^1=&\varphi_{02}''(h_{0,1}+h_{0,2})+\gamma_{2v} q^2\\
    E_2^2=&\varphi_{11}''(h_{0,2})+ \varphi_{02}''(h_{0,1}+h_{0,2})+ \gamma_{2,v}q^2.
\end{split}
\end{align}
To obtain a cleaner notation, we collect the prefactor-matrix into $A:= -q^2 M E$. Eq.~\eqref{eq:linear_thin_film_equation} can be easily solved to be 
\begin{align}
\label{eq:linear_solution}
    \begin{pmatrix}
        \widehat{\ h_1} \\ \widehat{h_2} 
    \end{pmatrix}(q,t)=\exp(A(q)t)\begin{pmatrix}
        \widehat{\ h_1} \\ \widehat{h_2} 
    \end{pmatrix}(q,0). 
\end{align}

We initialize simulations of two liquid layers with constant thickness and a small perturbation $h_i=h_{0,i}+\epsilon \mathcal{N}$. Here $\epsilon \ll h_i$ and $\mathcal{N}$ is a random variable. We take $\widehat{h_i}(q,0)$ to be its average value at the initial condition and compare  Eq.~\eqref{eq:linear_solution} with the Fourier transform of our simulations at different time steps. An example of such a simulation is shown in Fig.~\ref{fig:lsa}, reporting both thicknesses at two different time steps. The time steps are normalized by the rupture time $t_r$, i.e., the first time-step where one of the two layers (in the case of Fig.~\ref{fig:lsa} it is the upper layer) exhibits the formation of a hole in the performed simulation. Another possible choice for the time scale would be to take the time associated with the fastest growing mode as given by the LSA $\tau=1/\max\eta$, where $\eta_{\max}$ is the biggest eigenvalue of the matrix $A$, or as done by \cite{pototsky_alternative_2004} to consider the upmost layer as a single layer and derive the time scale as the time scale of the single layer film. We report one time step close to the initial condition, where only small perturbations of the liquid layers are present, and another time step closer to film rupture where the perturbation of the film is clearly visible. 

A good agreement between the simulation and the linear solution can be observed initially. All perturbations with a wave number bigger than the cutoff wave number $q_0$ present in the initial condition decay rapidly in the early time steps. The cutoff wave number $q_0$ is defined to be the largest $q$ such that $\max\eta_1(q)>0$ where $\eta_1(q)$ is the biggest eigenvalue of $A(q)$. Perturbations of wave-number $q<q_0$ grow exponentially in agreement with the linear solution Eq.~\eqref{eq:linear_solution}. A maximum in the spectrum of the thicknesses is formed at the fastest growing mode with wave-number $q_{\max}=\text{argmax}\eta_1(q)$. Upon growing, the perturbations become larger, and higher-order terms start influencing the evolution, leading to a deviation of the spectrum obtained by numerical integration. This effect can first be seen at large wave numbers. After film rupture, second-order terms dominate the dynamics of the system. The exponential growth of perturbations is halted, and the film evolves towards a steady state. After film rupture, the linear-solution Eq.~\eqref{eq:linear_solution} cannot describe the system's evolution anymore as higher-order terms dominate the evolution.

Performing simulations as the one shown in Fig.~\ref{fig:lsa} and comparing to the approximate linear solution Eq.~\eqref{eq:linear_solution} validates our numerical algorithm and shows that the LSA is a powerful tool for the study of multilayer systems. 

\subsection{Contact angle of a droplet}
\label{sec:angle}

\begin{figure}
    \includegraphics[width=\linewidth]{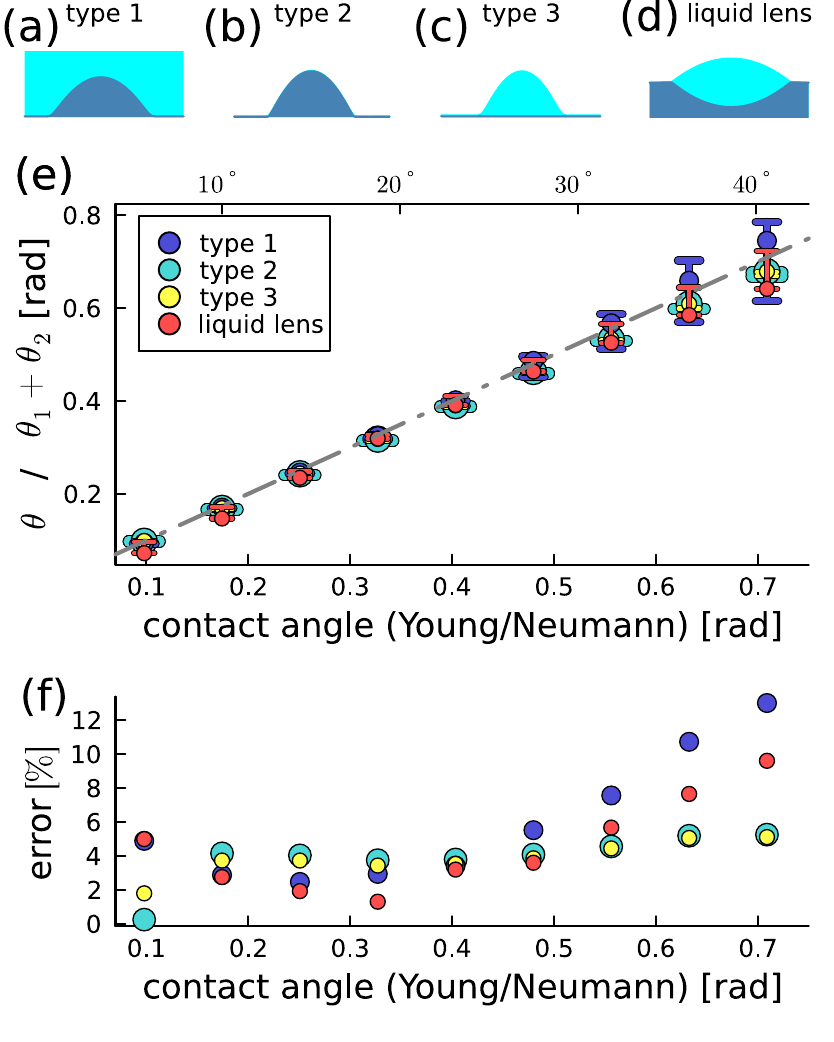}
    \caption{Relaxation of droplets to their expected equilibrium shape given by Young's law Eq.~\eqref{eq:youngs_law} or Neumann's law Eq.~\eqref{eq:neumanns_law}, respectively. (a) Type 1 droplet: a droplet of liquid one emersed in liquid two. (b) Typ 2 droplet: A droplet of liquid one in contact with the gaseous atmosphere.  (c) Type 3 droplet: A droplet of liquid two in contact with the gaseous atmosphere, resting on the solid substrate.  (d) Liquid lens: a lens-like droplet of liquid two in contact with the gaseous atmosphere floating on a bath of liquid one. (e) Simulations of droplets relaxing to their equilibrium contact angle showing expected vs. measured contact angle. For each data point, we report simulations with initially different contact angles, giving the error bars. (f) The maximum relative errors reported for all considered types of droplets. They are calculated by the absolute volume of the difference between the expected droplet shape and the simulated droplet shape divided by the total volume of the droplet.}
   \label{fig:relax}
\end{figure}

In this section, we study the equilibrium shape of individual droplets obtained from our model and compare them to the theoretically expected shapes. We need to consider two classes of droplets: \emph{sessile droplets}, i.e., drops of liquids resting on a solid substrate, and \emph{liquid lenses}, i.e., droplets of one liquid floating on another liquid and partly sinking into it. In the following, we will restrict ourselves to two liquid layers. 

For sessile droplets of two-liquid systems, we have to consider three types: a droplet of liquid one submerged in an atmosphere of liquid two (from now on called \emph{type 1}, see Fig.~\ref{fig:relax}(a)), a droplet of liquid one in contact with the gas atmosphere (called \emph{type 2}, see Fig.~\ref{fig:relax}(b)), and a droplet of liquid two resting on the bare solid substrate (called \emph{type 3}, see Fig.~\ref{fig:relax}(c)). Within our model, type 2 droplets rest on a precursor film of liquid one, and type 3 droplets are covered in a precursor layer of liquid two. In equilibrium, all three droplet types take the shape of a spherical cap whose contact angle $\theta$ (also called Young's angle) is given by \emph{Young's-law of wetting} \cite{doi_soft_2013, liang_surface_2018}
\begin{align}
\label{eq:youngs_law}
    \cos \theta = \frac{\gamma_{sv}-\gamma_{sl}}{\gamma_{lv}}.
\end{align}
Here, $\gamma_{sv}$, $\gamma_{sl}$, and $\gamma_{lv}$ refer to the solid-atmosphere (vapor), solid-liquid, and liquid-atmosphere surface tension. For sessile droplets of type 1 we have $\gamma_{sv}=\gamma_{02},\gamma_{sl}=\gamma_{01}, \gamma_{lv}=\gamma_{12}$, for the type 2 $\gamma_{sv}=\gamma_{03},\gamma_{sl}=\gamma_{01}, \gamma_{lv}=\gamma_{13}$, and finally for type 3 $\gamma_{sv}=\gamma_{03},\gamma_{sl}=\gamma_{02}, \gamma_{lv}=\gamma_{23}$.

A liquid lens is shaped as the intersection of two circles, consequently exhibiting two contact angles $\theta_1$ at the liquid two - atmosphere and $\theta_2$ at the liquid one - liquid two interface. Those so-called Neumann angles have to fulfill \emph{Neumann's law}
\cite{hack_self-similar_2020,scheel2023viscous,liang_surface_2018}
\begin{subequations}
\label{eq:neumanns_law}
    \begin{align}
        \gamma_{23}\cos\theta_1 +\gamma_{12}\cos\theta_2=&\gamma_{13}\\
        \gamma_{23}\sin\theta_1=&\gamma_{12}\sin\theta_2. 
    \end{align}
\end{subequations}

Exemplary simulation snapshots (not to aspect ratio) of all described kinds of droplets can be seen in the top panel of Fig.~\ref{fig:relax}.

It has to be mentioned that in experiments, Young's Eq.~\eqref{eq:youngs_law} and Neumann's law Eq.~\eqref{eq:neumanns_law} are not perfectly fulfilled. Instead, the measured equilibrium contact angle will be in a certain interval around the theoretical value called the \emph{static contact angle hysteresis}. This effect is associated with contact line pining caused by surface imperfections and accumulated charges in the liquid and on the substrate~\cite{butt_physics_2003, de_gennes_capillarity_2004}.

To test the model presented here, we simulate the relaxation of all kinds of droplets (type 1,2,3, liquid lens, see Fig.~\ref{fig:relax}(a)-(d)) toward their equilibrium. To do so, we vary the surface tensions such that by Young's law or Neumann's law, we expect contact angles between $0$ and $\pi/4=45^\circ$. We do not test the model beyond $\pi/4$ as this contact angle already represents $h(x)= x$ at the contact line. Thus, the assumption of length scale separation or lubrication approximation used to derive the model is not fulfilled. Furthermore, our numerical method is unstable for the used parameters beyond contact angles of $\pi/4$. For the liquid lenses, we restrict ourselves to top-down symmetric droplets by choosing $\gamma_{12}=\gamma_{23}$. It is, of course, possible to simulate asymmetric lenses by choosing $\gamma_{12}\neq\gamma_{23}$. The initial conditions used in our simulations are always droplets of the tested type obtained as intersections of a circle with the substrate for sessile droplets or the intersection of two circles for liquid lenses. We perform simulations starting from three different contact angles for every kind of droplet and test the value of the expected equilibrium contact angle.  One simulation starts at the expected contact angle (dots), and the other two start with droplets at twice or half the expected droplet angle (error bars). 

From visual inspection of the simulations, we observe the following general dynamics. An example of three simulations showing the relaxation of differently initialized liquid lenses can be seen in the supporting video \emph{relax\_video.mp4}. In this video, we report length scales in units of the maximum height $h_{\max}$ of the equilibrated droplet (i.e., the last simulation snapshot), and time in units of the time-scale $\tau=L\mu_2/\gamma_{23}$ where $L$ is the simulation box size, $\mu_2$ the viscosity of the liquid lens, and $\gamma_{23}$ the surface tension between the two liquids. Droplets relaxing by advancing the contact line, i.e., starting from a contact angle larger than expected, relax faster than droplets relaxing by withdrawing their contact line, i.e., starting from a contact angle smaller than expected. When advancing, the contact line is fast pushed outwards at a low apparent contact angle while the bulk of the droplet changes its curvature on a longer timescale, relaxing the drop to a spherical shape again. When withdrawing, the contact line moves slower and at a high local contact angle. This dynamics is in agreement with the dynamics seen from full hydrodynamics simulation~\cite{reddy_statics_2020, cheng_numerical_2018, craster_dynamics_2009}.


Starting simulations from a too-high or two-low contact angle, we observe that droplets never fully relax to the same value. Still, there is a difference between the relaxed simulated contact angle of a droplet starting from above or below the expected contact angle. The contact line gets pinned at some point close to the simulation's expected equilibrium. This ``numerical hysteresis'' reminds of the static contact angle hysteresis observed in experiments but is not of the same physical origin and is not to be taken too literally.  

Fig.~\ref{fig:relax} reports the results of the performed test described above. In inset (e), we show the measured contact angles from the numerical test described in the previous paragraph. Contact angles are measured by fitting a spherical arc to the droplet shape obtained from the simulation. To do so, we minimize the percentual volume difference  $E(\theta)=\int |h(x)-H_\theta(x)| dx / \int h(x) dx$ in $\theta$. $H_\theta(x)$ represents a spherical arc that shares the same peak position and value as $h(x)$ and has a contact angle $\theta$. The best fit to the simulation data exhibits errors smaller than $2\%$ or better in all simulations, indicating droplets are in good agreement with the expected spherical cap shape. We cross-checked the contact angles measured this way by measuring the droplets' width and height and calculating the contact angles from those values, as done in Ref.~\cite{zitz_lattice_2019}, with good agreement. We report the contact angle observed by initializing the simulation with the expected contact angle as circles and the values obtained by starting the simulations from above or below (the numerical contact angle hysteresis) by upper and lower error bars. We can see a generally good agreement between the measured values and those expected from Young's or Neumann's law, respectively. The agreement worsens upon approaching a contact angle of $\pi/4=45^\circ$. Droplets of type 1 show good agreement with Young's law for values up to $\pi/7\approx 25^\circ$. At larger expected contact angles, the droplet advancing from below shows strong numerical hysteresis. Type 2 and 3 droplets behave almost identically, showing good agreement and little numerical hysteresis for all tested values. Even though in good qualitative agreement with Neumann's law, the liquid lenses are simulated with too small opening angles for small expected angles. At an opening angle around $\pi/9=20^\circ$ the liquid lenses are in good agreement with Neumann's law. The simulated angles, again, are smaller than the expected angles for large opening angles. Here, numerical hysteresis is especially strong for simulations approaching the expected angle from above.

To make the discussion more precise, we report the relative deviation from the spherical cap with a contact angle predicted by Young's or Neumann's  law in Fig.~\ref{fig:relax}(f). The error reported in Fig.~\ref{fig:relax}(f) is calculated as the percentual volume difference  $E(\theta_0)=\int |h(x)-H_{\theta_0}(x)| dx / \int h(x) dx$, with $\theta_0$ the expected contact angle and $H_{\theta_{0}}$ a spherical arc with that contact angle. Out of the three performed simulations per droplet kind and expected contact angle, we report the one with the largest error. The error is always smaller than $15\%$, for many simulations even smaller than $6\%$. The largest errors are measured for type 1 droplets with contact angles larger than $\pi/7\approx 20^\circ$. 

To conclude this section, it can be said that Young's and Neumann's laws are well reproduced by our method. As expected, the model is more accurate for small angles.  According to this study, the best point of operation for our method is for $\theta_0<\pi/7$. 

\subsection{Dynamics of a relaxing corrugated upper layer}
\label{sec:dynamics}

\begin{figure}
\includegraphics[width=\linewidth]{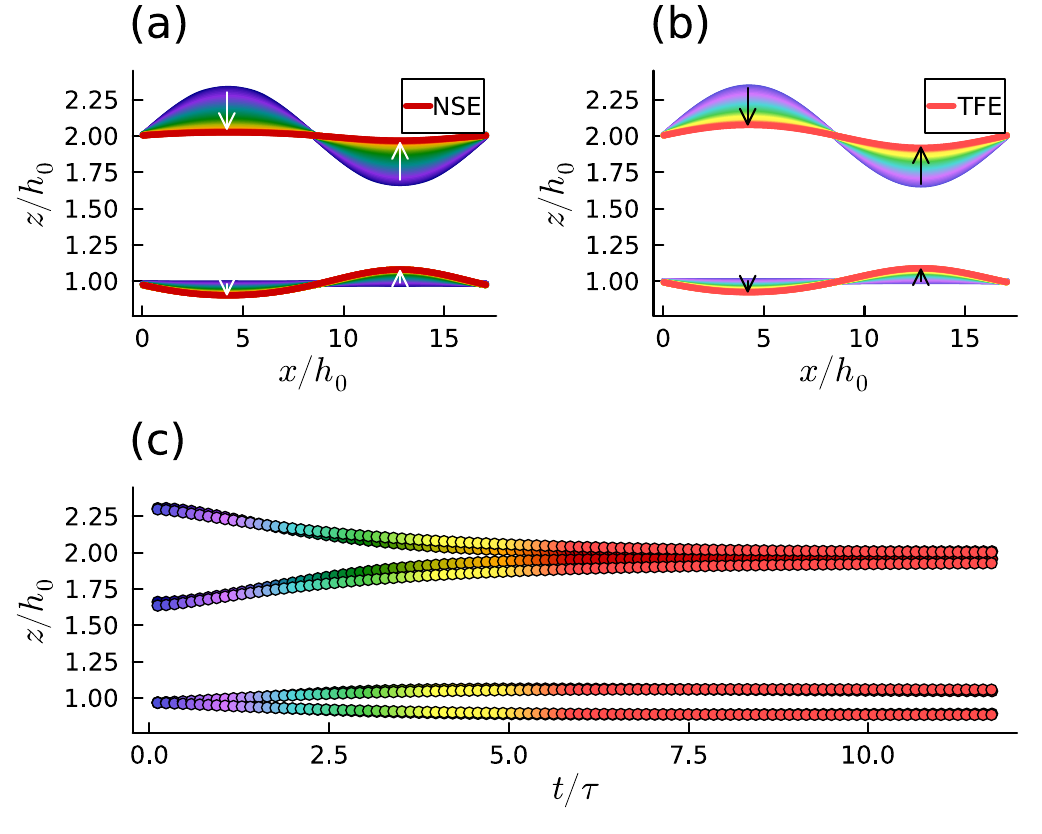}
\caption{Comparing the simulation of a relaxing sine-wave in a two-layer film between full Navier-Stokes-equation (NSE) and thin film equation (TFE) simulations. Lengths are reported in units of the mean film thickness $h_0$ and times in terms of the visco-capillary time $\tau=\frac{\mu L}{\gamma}$ where $L$ is the initial wavelength of the sine wave. Subplot (a) refers to the NSE simulation, and subplot (b) refers to the TFE simulation. In both plots, the lower set of lines shows the position of the interface between liquid one and liquid two $z_1$, while the upper set of lines shows the position of the interface between liquid two and the gas atmosphere $z_2$. The colour code refers to the time step and can be read off of the $x$-axis of subplot (c). Early-time steps are reported in blue, and late-time steps in red, transitioning via green and yellow. Data points from the NSE simulation are darker than those from the TFE simulation. Subplot (c) reports (in dark colours for the NSE simulation and light colours for the TFE simulation) the maximum of the upper interface $\max z_2$ over time as the upmost set of points. The minimum of the upper interface $\min z_2$ is the second upmost set of points, while the lowest two sets of points refer to the maximum and minimum of the lower interface $\max z_1, \min z_1$.}
\label{fig:dynamics}
\end{figure}

In order to assess the accuracy of the dynamics of the here presented model, we compare numerical results obtained by the method described in Sec.~\ref{sec:LBM} with the numerical solution of the full Navier-Stokes-equation (NSE). Those full NSE simulations are carried out via an LBM-color-gradient algorithm~\cite{leclaire_generalized_2017} in two dimensions using the same implementation as employed by~\cite{scheel2023viscous, scheel2024enhancement}. We initialize the simulations with the lower layer being at constant height $h_1(x)\equiv h_0$ while the upper layer we initialize as a sine-wave $h_2(x)=h_0+\delta h_2 \sin( 2\pi x/L)$, where $L$ is the horizontal size of the simulation box. That is, we fit a single period of the sine-wave into the simulation box. In the $x$ direction, we choose periodic boundary conditions. In the $z$ direction, we choose a wall boundary for the full NSE simulations, while the TFE simulation does not have a boundary in that direction. We choose surface tensions $\gamma_{1,2}=\gamma_{2,3}$ and $\mu_1=\mu_2$. In the TFE simulation, we implicitly have $\mu_3=0$, while for the full NSE simulation, we choose  $\mu_3=\mu_1/10$. The spreading parameters we choose as $S_{01}=S_{11}=S_{02}=0$ such that the disjoining pressure is not relevant in the TFE simulation. The disjoining pressure is not implemented in the full NSE simulation. 

The results of our simulations are presented in Fig.~\ref{fig:dynamics}. Fig.~\ref{fig:dynamics}(a) displays the  time evolution of the two interfaces $z_1$ and $z_2$ obtained from the full NSE simulation while Fig.~\ref{fig:dynamics}(b) displays the time evolution of the two interfaces obtained from the TFE simulations. Both simulations show that the initial sine-wave of the upper $z_2$ interface flattens out over time. Due to the positive curvature of the $z_2$ interface in the left half of the simulation box, the pressure jump between the atmosphere and the layer of liquid two $p_2-p_{atm}$ is higher, as compared to the right half of the simulation box where the curvature of the $z_2$ interface is negative. This leads to a decrease in film height $z_2$ on the left and an increase of the film height $z_2$ on the right. This higher or lower pressure in the upper liquid layer further leads to higher or lower pressure, respectively, in the lower liquid layer. Consequently, the lower interface $z_1$ is decreased on the left and increased on the right. We observe the initially flat interface $z_1$ to become corrugated due to the initial corrugation of the upper interface $z_2$. As the upper interface $z_2$ is in contact with the non-viscous gas atmosphere, its dynamics are faster and less damped than the dynamics of the lower interface $z_1$ that is in contact with the viscous liquids one and two. We expect the lower interface $z_1$ to relax and be flat again on time scales much longer than the ones simulated here.

In general, a good agreement between the NSE simulation (Fig.~\ref{fig:dynamics}(a)) and the TFE simulation (Fig.~\ref{fig:dynamics}(b)) can be seen from those plots. To better compare the two simulations, we show in Fig.~\ref{fig:dynamics}~(c) the time evolution of the maxima and minima of the two respective interfaces. One can see a good agreement between the two data sets, even though it has to be said that in the full NSE simulation, the upper interface $z_2$ evolves slightly faster. As time evolves, we observe how the extrema of the upper interface  $\max z_2$ and $\min z_2$ align while the extrema of  the lower interface $\max z_1$ and $\min z_2$ deviate from each other. 

We finalize this section by commenting on simulation times. The full Navier Stokes simulation shown in Fig.~\ref{fig:dynamics}(a) required a total of $105 \text{CPU}\cdot\text{h}$ on 128 \emph{AMD EPYC 7662} processor-cores taking a wall-clock-time of almost 50 minutes. It must be mentioned that the implementation used here is three-dimensional, with one direction chosen negligibly small (4 lattice nodes), making it not performance optimal for two-dimensional simulations. On the other hand, our implementation of the multilayer TFE performed the simulation shown in Fig.~\ref{fig:dynamics}(b) on a single core in less than two minutes, taking only $0.03\text{CPU}\cdot\text{h}$. This leaves a factor of 3400 between the performance of the two simulations, rendering the full NSE simulation infeasible without large computing facilities. In contrast, the TFE simulation can be easily run on a single workstation. This considerable speed-up of the TFE simulation method compared to the full NSE method stems from reducing the problem's dimensionality, solving a $(d-1)$-dimensional PDE for a $d$-dimensional situation.

\subsection{Equilibrium states of thin two-layer films}
\label{sec:configs}

\begin{figure}
    \centering
    \includegraphics[width=\linewidth]{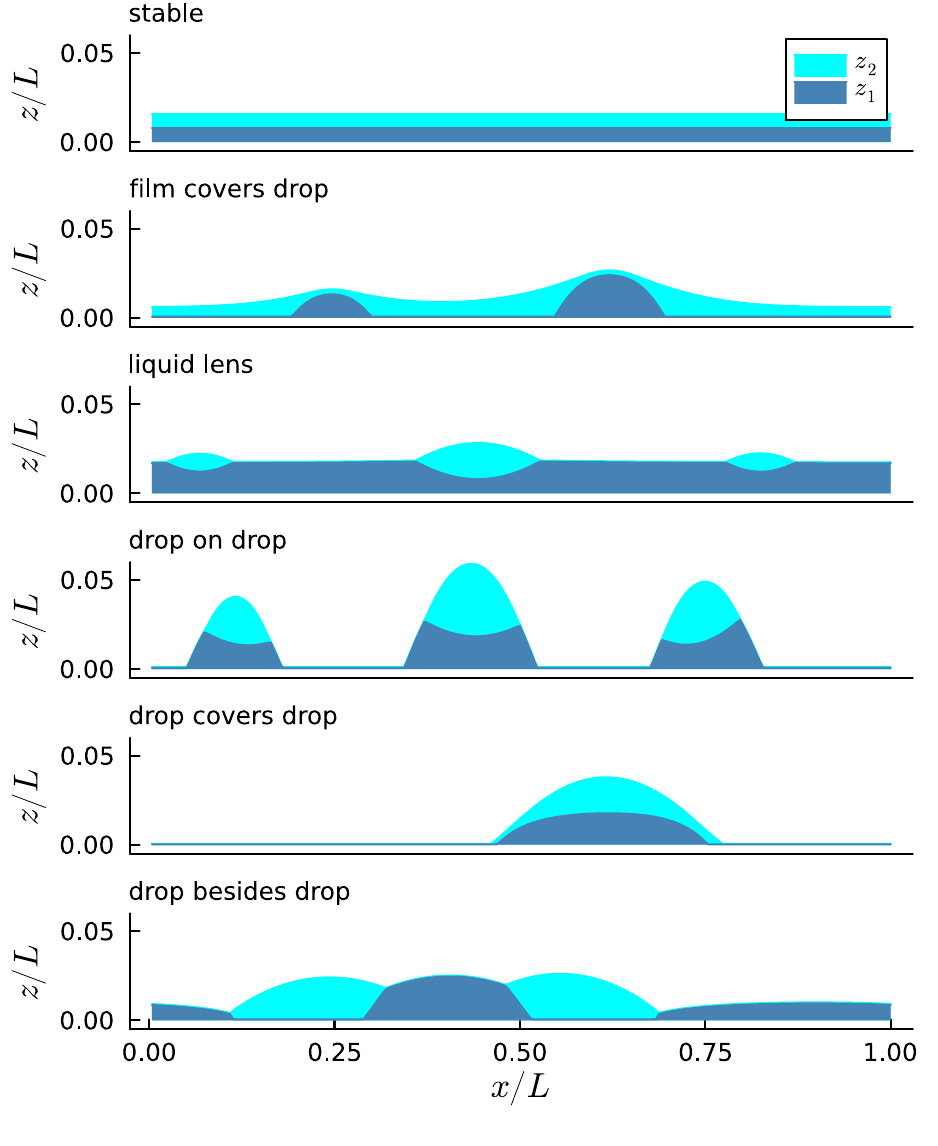}
    \caption{Simulation snapshots with varying spreading coefficients $S_{01},S_{11},S_{02}$ (as described in the main text) showing different possible configurations after film rupture. All length scales are reported in units of the simulation box size $L$.}
    \label{fig:configs}
\end{figure}

This subsection showcases the different configurations a film of two liquid layers can attain after film rupture as described by Ref.~\cite{diekmann_mesoscopic_2024}, from which we adopt the nomenclature. We initialize the two liquid layers with constant height, and we add a small perturbation $h_i=h_{0,i}+\epsilon\mathcal{N}$ as already done in subsection \ref{sec:lsa}. To showcase the different possible configurations, we vary the total volume of the two liquid layers and thus choose different values of $h_{0,1}$ and $h_{0,2}$ across different simulations and vary the spreading parameters $S_{01},S_{11},S_{02}$, resulting in the different configurations reported in Fig.~\ref{fig:configs}. A negative spreading coefficient creates a disjoining pressure driving a liquid layer apart, ultimately leading to film rupture. A spreading coefficient of zero renders the disjoining pressure ineffective. Finally, a positive spreading coefficient stabilizes a liquid layer, hindering film rupture. The spreading parameters $S_{01}$, $S_{11}$, and $S_{02}$ correspond to the lower layer, upper layer, and combined disjoining pressure, respectively.  It has to be mentioned that even for a large spreading coefficient, the combined disjoining pressure is typically weak as long as none of the layers is dewetted because the thickness of both layers combined is trivially larger than the thickness of a single layer. The simulations in Fig.~\ref{fig:configs} depict intermediate configurations rather than the final equilibrium, which may take a long time to reach. However, these snapshots provide insight into the characteristic appearance of the eventual equilibrium state.

In the following, we describe the choices of spreading coefficients leading to the configuration shown in the simulation snapshots shown in Fig.~\ref{fig:configs}. 
Choosing $S_{01}=S_{11}=S_{02}\geq 0$, the disjoining pressure is not active ($=0$) or hindering film breakage ($>0$), while the Laplace pressure is suppressing corrugation of the liquid interfaces. Consequently, the homogenous system is a \emph{stable} steady state.

Choosing $S_{01}< 0$ creates a disjoining pressure acting on the lower liquid layer, only breaking it into droplets. Having  $S_{11}>0$, in the meantime, stabilizes the upper layer. Furthermore, setting $-S_{02}<S_{11}$ ensures that the upper layer is not broken by the disjoining pressure stemming from the attraction of the surrounding gas atmosphere and the solid support. This attraction becomes important once the lower liquid layer is broken. One observes a \emph{film covers drop} state as shown in the second subplot of Fig.~\ref{fig:configs}. 

For $S_{01}> 0$, the lower layer is stabilized, while $S_{11}$ leads to a disjoining pressure stemming from the liquid 1 gas attraction that breaks the upper liquid layer into so-called \emph{liquid lenses}, as can be seen in the third panel of Fig.~\ref{fig:configs}. $S_{02}$ is chosen such that $-S_{02}<S_{01}$, ensuring the lower film is not broken by the associated disjoining pressure after the breakup of the upper layer. 

For a choice of spreading parameters with all three negative $S_{01}, S_{11}, S_{02}<0$, the associated disjoining pressures will try to minimize regions wetted by liquid 1, regions wetted by liquid 2, and regions wetted by both liquids at once. The result is the formation of a liquid lens of the upper liquid 2 sitting on a drop of the lower liquid 1. Such a \emph{drop on drop} state can be seen in the fourth subplot of Fig.~\ref{fig:configs}.

Choosing $S_{01}<0$, the lower layer breaks. Then choosing $S_{11}>0$ and $0>-S_{11}>S_{02}$ leads to the upper liquid layer being stable when in places where the lower layer exists but being unstable due to the disjoining pressure acting on the combined layer thickness wherever the lower layer is already dewetted. This leads to the upper liquid breaking and forming a contact line in regions with no droplets of the lower liquid while keeping a closed film over the droplets of the lower liquid. We observe \emph{drop covers drop} shapes as shown in the fifth subplot of Fig.~\ref{fig:configs}.

When $S_{01}, S_{11}<0$, both layers tend to break and form droplets. Choosing $S_{02}>0$, it is energetically unfavourable to uncover the solid substrate, leading to droplets of liquid 1 and 2 being placed next to each other in a \emph{drop besides drop} state as displayed in the last panel of Fig.~\ref{fig:configs}. Upon fine-tuning the spreading coefficient and the total volumes of the two liquids, one can switch between cases exposing some of the solid substrate or not exposing it. 

\onecolumngrid
\begin{center}
\begin{figure}[h]
    \includegraphics[width=\linewidth]{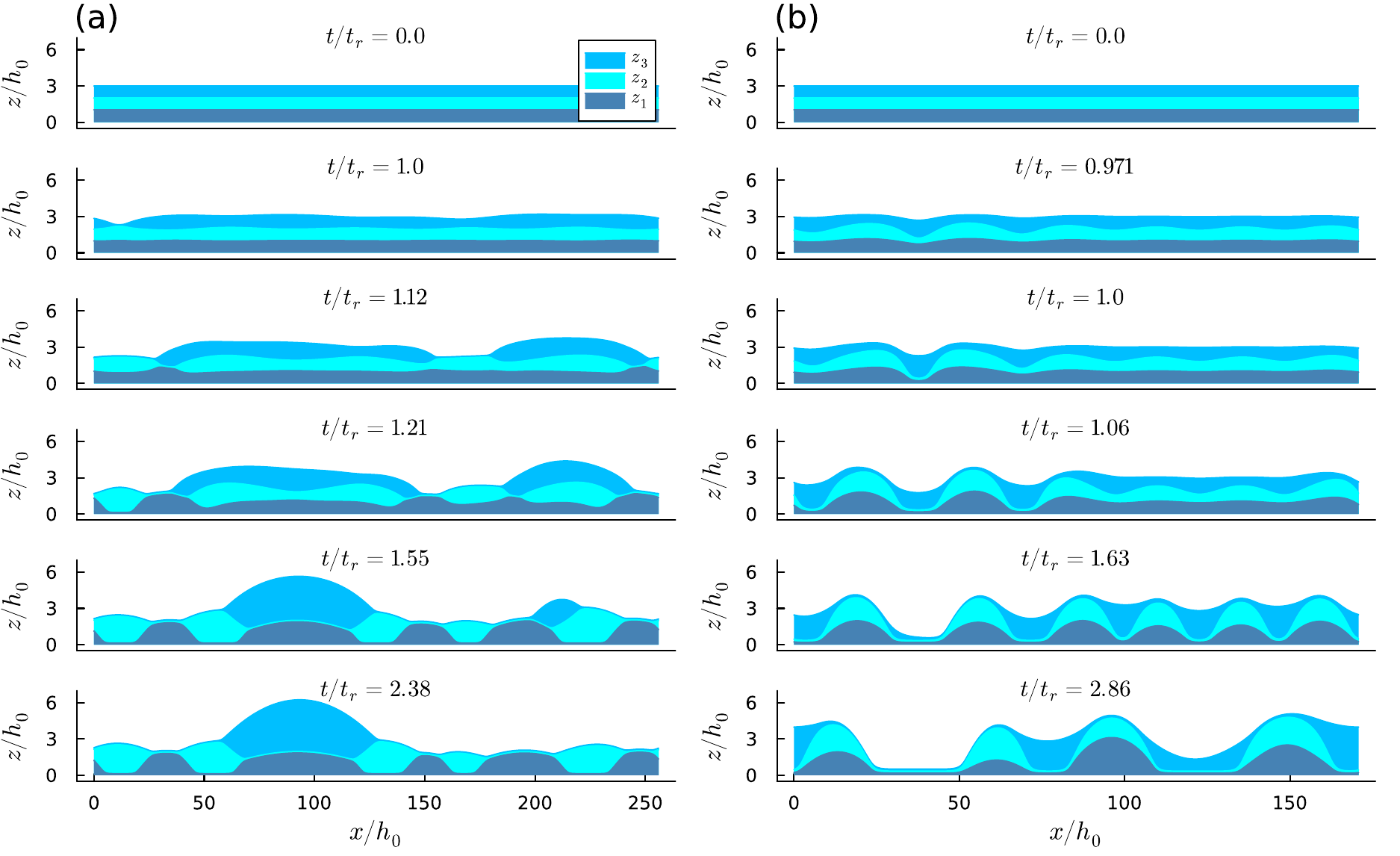}
    \caption{Simulation snapshots of a three-layer thin film undergoing spinodal rupture. The time is reported in units of the rupture time $t_r$ measured from the simulation. All length scales are reported in units of the initial film thickness $h_0$. The two shown simulations differ by the chosen spreading coefficients: in (a) we choose $S_{01}=S_{11}=S_{21}<0$ and $S_{02}=S_{12}=S_{03}=0$ resulting in a varicose mode while in (b) we choose $S_{01}=S_{11}=S_{21}<0$, $S_{02}<S_{01}$ and $S_{12}=S_{03}=0$ resulting in a zigzag mode. }
    \label{fig:3layers}
\end{figure}
\end{center}
\twocolumngrid

\subsection{Spinodal rupture of three layers}
\label{sec:3_layers}

After discussing thin films consisting of two liquid layers, we use our approach to simulate a thin film system consisting of the liquid layers undergoing spinodal rupture driven by the disjoining pressure. To do so, we prepare a numerical simulation, solving Eq.~\eqref{eq:thin_film_equation_3} by the method introduced in Sec.~\ref{sec:LBM}, in a homogenous initial state with a small perturbation. That is, we initialize the system as $h_i=h_0+\epsilon \mathcal{N}$ for $i=1,2,3$, $h_0$ a constant value equal for all three indices, $\epsilon\ll h_0$, and $\mathcal{N}$ a random variable. Furthermore, we choose all spreading coefficients $S_\xi<0$ for $\xi = 01, 11, 12$, i.e., we assume all layers to be prone to disjoining, and $S_\xi=0$ for $\xi=02, 03, 12$, i.e., no disjoining pressure acting on multiple layers combined.  In addition, we choose the viscosities equal for all three liquids $\mu_1=\mu_2=\mu_2$, and the same for the surface tensions with the above phase $\gamma{12}=\gamma_{23}=\gamma_{34}$. Parameters are chosen such that viscous forces dominate. Snapshots of this simulation are shown in Fig.~\ref{fig:3layers}(a). We report the time in units of $t_r$, i.e., the first step in which a hole has nucleated in one of the layers as measured by simulation. A second simulation of the spinodal dewetting of three liquid layers is shown in Fig.~\ref{fig:3layers}(b). This simulation uses the same parameters as the one in Fig.~\ref{fig:3layers}(a) with the exception of the spreading parameter $S_{02}$, which is chosen $S_{02}<S_{01}$ in the second simulation.

We first describe the evolution of the simulation shown in Fig.~\ref{fig:3layers}(a). 
The first layer to break (at $t/t_r=1$ by definition) is the upmost layer of liquid three, as the evolution of this one is least slowed down by the friction with the solid substrate. One can observe that the $z_3$ interface is decreasing in height wherever the $z_2$ interface is increasing and vice-versa. This phenomenon is termed a \emph{varicose} mode in the nomenclature of~\cite{pototsky_alternative_2004,pototsky_morphology_2005, pototsky_evolution_2006}, contrasting with a \emph{zigzag} mode where both interfaces move in parallel, as seen in Fig.~\ref{fig:3layers}(b).

While liquid three contracts to droplets, the mid-layer of liquid two breaks at $t/t_r=1.12$. The attraction between liquid one and liquid three pierces a hole into liquid two, bringing the two liquids in contact. This again happens in a varicose mode. 

Finally, at $t/t_r=1.21$, the lowest layer of liquid one breaks in a varicose mode, bringing droplets of liquid two in contact with the solid substrate.

Running the simulation further, we see at $t/t_r=1.55$ a layer of droplets of liquid one with droplets of liquid two in between, completely covering the substrate. Given the total volumes of liquid one and two, our choice of $S_{02}=0$, there is no reason to expect the solid substrate to be exposed to the gas atmosphere upon running the simulation longer. The two lower liquids are in a \emph{drop besides drop} state as described in section \ref{sec:configs}. On top of this, liquid three rests in  two large droplets. At time step $t/t_r=2.38$, the smaller one of the two droplets of liquid three was absorbed by the bigger droplet. The transport occurs via the thin precursor layer and is driven by the pressure gradient due to the larger curvature of the smaller droplet. 

The simulation is not yet equilibrated at the final reported time step. At equilibrium, we expect only a single droplet of each liquid, i.e., the droplets of liquid one and two, to unify in the same way we have seen for liquid three or due to droplet coalescence. 

In the simulation shown in Fig.~\ref{fig:3layers}(b), we include a strong attraction between liquid three and the solid substrate by choosing $S_{02}<S_{01}=S_{11}=S_{21}$. Due to the extra contribution to the disjoining pressure stemming from this choice, film rupture occurs first for layers one and two, which break simultaneously at $t/t_r=1$. This film breakup happens in a zigzag mode, meaning that the thicknesses of layers one and two are approximately proportional to each other during the dynamics leading to film rupture. 

Shortly after the breakup of the first two layers, the third layer breaks at the maxima of the underlying interface $z_2$, exposing liquid two to the gas atmosphere ($t/t_r=1.06$). 

While droplets of liquid one and two in the drop covers drop state (compare section~\ref{sec:configs}) form, liquid three accumulates in the grooves between those droplets. This is expected as we have modelled a low surface energy, or, said differently, a strong attraction between liquid three and the solid substrate. 

At $t/t_r=1.63$ we see the first exposure of the solid substrate to the gas atmosphere when liquid three dewets on a part of the substrate that has already been dewetted by liquids one and two. When the simulation proceeds, this deweted region grows, and the system coarsens by droplet coalescence. Again, we do not run the simulation fully to equilibrium, where we expect further coarsening and only a single droplet per liquid.


\onecolumngrid
\begin{center}
\begin{figure}[h]
\includegraphics[width=\linewidth]{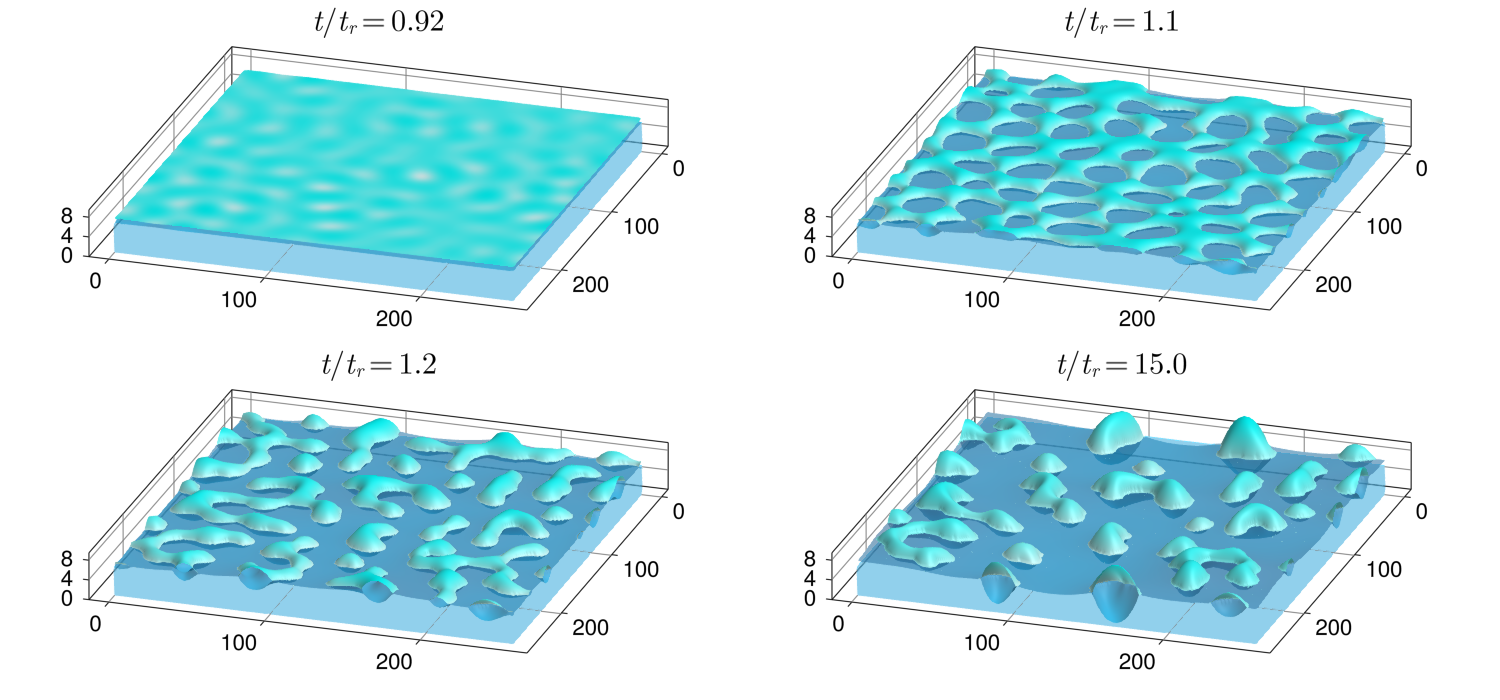}
\caption{Spinodal rupture of a thin liquid layer resting on a thicker liquid layer simulated in 3D. Axes are in units of the initial height of the upper layer $h_{0,2}$. Time is reported in units of the rupture time $t_r$ measured from the simulation.}
\label{fig:3D}
\end{figure}
\end{center}
\twocolumngrid

\subsection{Spinodal rupture of two layers in 3D}
\label{sec:3d}

Finally, we present a pseudo-three-dimensional simulation of a thin liquid layer's spinodal rupture resting on a thicker layer, using two-dimensional height fields $h_1$ and $h_2$. Such simulations have also been presented and discussed in Ref.~\cite{pototsky_evolution_2006}. We do not discuss this topic here exhaustively but only showcase the ability of our numerical algorithm, as described in Sec.~\ref{sec:LBM}, to simulate large three-dimensional systems. As our method is an LBM, parallelization is straightforward. For two-dimensional fields, simulation time can be significantly improved by using GPUs as described by Zitz et al.~\cite{zitz_lattice_2019,zitz_swalbejl_2022}. With our implementation the simulation presented in Fig.~\ref{fig:3D} can be performed in $95$ minutes on a \emph{NVIDIA Titan V} GPU. 

We initialize the system with constant height values and small perturbations $h_i=h_{0,i}+\epsilon \mathcal{N}$ for two constants $h_{0,1},h_{0,2}$ where $h_{0,1}=6h_{0,2}$, $\epsilon\ll h_{0,2}$, and $\mathcal{N}$ is a random variable. For the spreading coefficients, we choose $S_{01}>0$, $S_{02}<0$, and $S_{02}=0$, rendering the upper layer unstable due to disjoining pressure and the lower layer stable. The viscosities are $\mu_1=\mu_2$ and the surface tensions of the two liquids with the above phases are also chosen equal $\gamma_{12}=\gamma_{23}$. We report snapshots of this simulation in Fig.~\ref{fig:3D}. The time is reported in units of the rupture time. The precursor layer of liquid two is not shown for a clearer illustration. 

Due to the disjoining pressure acting on the upper layer, both layers start being corrugated ($t/t_r=0.92$). The lower liquid layer and the surrounding gas atmosphere attract each other, leading to an instability of the system that grows as a varicose mode. Once the film ruptures ($t/t_r=1.0$), a number of holes nucleate in the layer of the upper liquid two ($t/t_r=1.1$). As the evolution toward equilibrium continues, the rims separating the individual holes in the film break, resulting in a labyrinth structure of rims of liquid two ($t/t_r=1.2$). As more and more of the rims break, the structure ceases to percolate and approaches a state of single droplets distributed over the simulation domain. As time proceeds, the larger droplets will grow while the smaller ones shrink, either by coalescence or liquid transport within the precursor layer, driven by the higher pressure in the smaller and more curved droplets. The expected equilibrium state is a single liquid-lens of liquid two floating on liquid one. 

\section{Conclusions}
We have presented a novel thin film model for the hydrodynamics of arbitrarily many layers of immiscible liquid films stacked on top of each other in the lubrication approximation. Further, we have presented an LBM-based integration scheme to simulate such systems numerically. 

To showcase our method's abilities and test its physical accuracy, we performed and described a number of different simulations, including simulations of two and three liquid layers in two and three dimensions. We have shown that the numerical solver matches the expected linear solution from LSA and that different kinds of droplets, including sessile droplets and liquid lenses, attain the correct shape and contact angle after relaxation. Furthermore, we compared a simulation performed by our method with a similar simulation of the full hydrodynamics and then presented and briefly discussed the possible states that a system initialized as two homogeneous liquid layers can attain after potential rupture. We simulated and described the spinodal dewetting of three stacked liquid films. To the best of our knowledge, this is the first instance of such an endeavor. Finally, we showcased a three-dimensional simulation of the spinodal dewetting of a thin liquid film resting on a stable, thicker liquid film. 

Apart from the novelty of our model that accomplishes the step from the known thin-film-equations for one or two liquid layers, we want to highlight the effectiveness of our numerical algorithm that allows for the simulation of such systems on a single workstation without the need for high-performance computing facilities. Thanks to the ability of the model to reduce the effective dimension of the problem and  the algorithm to be operated without the need for re-meshing at the contact line, pseudo-two-dimensional simulations are easily set up and can be run serially on a single-core CPU. The LBM allows for easy parallelization on CUDA architectures, enabling fast pseudo-three-dimensional simulations. 

In the future, we plan to extend our model, including solutes and surfactants. We plan to further study the dynamics of systems including three or more layers and aim at technologically relevant applications, including coating and ink-jet printing techniques.

\subsection*{Data Availability}
The data necessary to recreate this manuscript is available under \url{doi.org/10.5281/zenodo.13828457}. 


\subsection*{Acknowledgments}
We acknowledge funding by the Deutsche Forschungsgemeinschaft (DFG, German Research Foundation)—Project-ID 431791331—SFB 1452. Furthermore, we acknowledge the Helmholtz Association of German Research Centers (HGF) and the Federal Ministry of Education and Research (BMBF), Germany, for supporting the Innovation Pool project “Solar $H_2$: Highly Pure and Compressed”. 

\section*{Apendix}

\subsection*{Free energy ansatz}

The pressure used in our model can be derived by the variation of a free energy functional. The free energy of the system can be formulated by generalizing the free energy of a single layer film \cite{thiele_gradient_2016, de_gennes_capillarity_2004}, as the area of each layer times the surface tension with the next higher layer plus a summed wetting potential $W_i$ modeling the wetting or dewetting of the layers above the $i$th layer:
\begin{align}
\label{eq:free_energy_n_layer_0}
\begin{split}
        \mathcal{F}=&\sum_{i=0}^{n} \int   \gamma_{i,i+1}\sqrt{1+|\nabla z_i|^2} + W_i(z_i, \cdots, z_n) d\mathbf{x}. 
\end{split}
\end{align}
The summed wetting potential $W_i$ introduces the interaction of all layers with indices $j>i+1$ with the $i$th layer. The interaction with the below layers is included as we sum over all $i$. We know that for a thick layer of liquid on top of the $i$th layer, there are no interactions of the $i$th with any above layer but the $i+1$th layer. Thus, 
\begin{align}
    \lim_{z_{i+1}-z_i\to \infty}W_{i}=0.
\end{align}
If all layers between the $i$th and $j$th layers dewet, the interface between fluid $i$ and fluid $i+1$ is replaced by the interface between fluid $i$ and fluid $j$ while all other interfaces in between do not exist. Consequently, the interactions among the layers above the $i$th layer and below the $j$th layer must vanish. We get
\begin{align}
   \begin{split}
        \lim_{z_{j}-z_i\to (j-i)h^*}W_i=&(\gamma_{ij}-\gamma_{i,i+1})\sqrt{1+|\nabla z_i|^2} \\&-\gamma_{i+1,j}\sqrt{1+|\nabla z_{i+1}|^2}.
   \end{split}
\end{align}
We recall that we are summing over all $i$ such that certain terms will cancel and that we use a precursor length $h^*$ in our modeling. This means that no layer will be thinner than $h_k=h^*$, considered a dewetted film. Finally, we choose the wetting potential $W_i$ such that it diverges when $z_j-z_i<(j-i)h^*$ for any $j>i$. As such, we do not allow values of $z_j-z_i<(j-i)h^*$. This penalizing contribution to the free energy does not affect the system's dynamics for thicknesses above the precursor length but assures the establishment of the precursor layer. One can, of course, choose $h^*=0$, but this will create divergence of the friction force upon dewetting any liquid layer. 

Having established the limits of the wetting potential $W_i$ we make an ansatz about its functional form. One might think that $W_i$ should vary over a molecular length scale around $h^*$, but integrating the van der Waals interaction between the individual layers shows that it is not so. One can show that the wetting potential $W_i$ decays for uncharged surfaces as $W_i \propto 1/(z_j-z_i)^2$ \cite{de_gennes_capillarity_2004, israelachvili_intermolecular_nodate}. For charged surfaces, it decays as $W_i\propto \exp(-\kappa (z_j-z_i))$ \cite{israelachvili_intermolecular_nodate}. This long-rangedness (up to $100\text{nm}$) of the wetting potential is what leads to the disjoining pressure, leading to film rupture for films with thicknesses up to $100\text{nm}$, which has been observed and described many times \cite{fetzer_thermal_2007, rauscher_spinodal_2008, vrij_possible_1966, oron_long-scale_1997, craster_dynamics_2006, renger_investigation_2000, shiri2021spinodal}. Due to our discussion on the wetting potential, we make the ansatz 
\begin{align}
    W_i=\sum_{k=1}^{n-i}\varphi_{ik}(z_{i+k}-z_i)\sqrt{1+|\nabla z_i|^2 },
\end{align}
which fulfills all the mentioned requirements. The function $\varphi$ is defined by Eq.~\eqref{eq:wetting_potential}. 

Applying the long-wave-approximation 
\begin{align}
    \sqrt{1+|\nabla z_i|^2}\approx 1 + \frac{(\nabla z_i)^2}{2}
\end{align}
that holds thanks to the lubrication approximation $(\nabla z_i)^2\ll 1$ we can finally write the free energy as
\begin{align}
\label{eq:free_energy_n_layer}
\begin{split}
        \mathcal{F}=&\sum_{i=0}^{n} \int   \gamma_{i,i+1}\left(1+\frac{(\nabla z_i)^2}{2}\right)\\
        &+\sum_{k=1}^{n-i}\varphi_{ik}(z_{i+k}-z_i)\left(1+\frac{(\nabla z_i)^2}{2} \right) d\mathbf{x}. 
\end{split}
\end{align}

From Eq.~\eqref{eq:free_energy_n_layer} the pressure jump across the $i$-$(i+1)$ interface can be obtained by calculating the variation
\begin{widetext}
    \begin{align}
        \begin{split}
        p_i-p_{i+1}&=\frac{\delta \mathcal{F}}{\delta z_i}= \underbrace{- \Bigg[ \gamma_{i,i+1} +\sum_{k=1}^{n-i} \varphi_{ik}(z_{i+k}-z_i)\Bigg] \nabla^2z_i}_{\text{Laplace pressure $\Lambda_i$}}\\
    &\underbrace{-\sum_{k=1}^{n-i} \varphi_{ik}'(z_{i+k}-z_i)\left( 1 +\frac{(\nabla z_i)^2}{2}\right)+ \sum_{l=0}^{i-1}\varphi_{l,i-l}'(z_i-z_l) \left( 1 +\frac{(\nabla z_l)^2}{2}\right)}_{\text{disjoining pressure $\Pi_i$}}. 
 \end{split}
    \end{align}
\end{widetext}
This result is in agreement with the pressure stated in the main text. \\
The multilayer thin film equation takes the form of a Cahn-Hilliard equation
\begin{align}
    \begin{pmatrix}
        h_1\\ \vdots \\ h_n 
    \end{pmatrix}=\nabla \cdot\left( M \nabla \begin{pmatrix}
        \frac{\delta \mathcal{F}}{\delta h_1}\\ \vdots \\ \frac{\delta \mathcal{F}}{\delta h_n}
    \end{pmatrix}\right).
\end{align}
Thus, the model considered in this paper is a conservation equation minimizing the free energy Eq.~\eqref{eq:free_energy_n_layer} under the mass conservation constraint. We know the physical mechanism driving the system to equilibrium as we derived Eq.~\eqref{eq:free_energy_n_layer} from surface tension and van der Waals attraction of distant layers.  

\subsection*{Limit cases of the two-layer model}

The height $h$ of a single-layer thin film can be described by the well-known single-layer TFE 
\begin{align}
\label{eq:thin_film_equation_1}
    \partial_t h = \nabla \cdot \left( \frac{h^3}{3\mu} \nabla p (h)\right) ,
\end{align}
where the pressure is given by 
\begin{align}
    p(h)=-\gamma\nabla^2 h + S\varphi'(h).
\end{align}
The spreading coefficient reads $S=\gamma_{sv}-\gamma_{sl}-\gamma$,  the surface tensions $\gamma, \gamma_{sv},\gamma_{sl}$ are associated to the  liquid-gas(vapor), solid-gas, and solid-liquid interfaces, and $\varphi=\varphi_{1}$ is the wetting potential as defined in Eq.~\eqref{eq:wetting_potential}. 

In this section, we demonstrate how the single-layer thin film equation Eq.~\eqref{eq:thin_film_equation_1} can be obtained from different limits of the two-layer thin film equation Eq.~\eqref{eq:thin_film_equation_2}. 

\subsubsection{$z_2\to \infty, \mu_2\to 0$}

In  the limit $z_2\to \infty, \mu_2\to 0$ the upper liquid layer becomes infinitely thick and non-viscous, taking the role of the gas atmosphere.

Bearing in mind that $\lim_{h\to\infty}\varphi'_k(h)=\lim_{h\to\infty}\varphi_k(h)=0$, we derive the limit of the pressure jumps to be 
\begin{subequations}
        \begin{align}
        \lim_{z_2\to\infty} (p_1-p_2)=&  \varphi_{01}'(z_1)- \gamma_{12}\nabla^2 z_1\\
        \lim_{z_2\to \infty}(p_2-p_{atm}) =& 0 .
        \end{align}
   \end{subequations}
With $\gamma_{01}=\gamma_{sl}$, $\gamma_{02}=\gamma_{sv}$, and $\gamma_{12}=\gamma$ we obtain $S_{01}=S$ and $\lim_{z_2\to\infty} (p_1-p_2)=p(h)$. 

Now we calculate the limit of the mobility $M$.
\begin{align}
     \lim_{\mu_2 \to \infty}\lim_{z_2\to \infty} M =&\begin{pmatrix}
        \frac{z_1^3}{3\mu_1} & \infty\\
        \infty & \infty
    \end{pmatrix}.
\end{align}
We thus have to calculate the limit of the product of the top right entry of the mobility $M_1^2$ with the pressure jump $p_2-p_{atm}$ explicitly. $p_2-p_{atm}=\mathcal{O}(\varphi')=\mathcal{O}(z_2^{-m-1})$ while $M_1^2=\mathcal{O}=z_2$   we obtain the limit to be $\lim_{\mu_2 \to \infty}\lim_{z_2\to \infty} M_1^2\nabla \nabla (p_2-p_{atm})=0$. Thus, the first line of the two-layer TFE Eq.~\eqref{eq:thin_film_equation_2} in this limit is indeed the single-layer TFE Eq.~\eqref{eq:thin_film_equation_1}. 

\subsubsection{$z_2\to z_1+h^*$}

The limit $z_2\to z_1+h^*$ can be equivalently formulated as $h_2\to h^*$, i.e., the thickness of the upper layer becoming as thick as the precursor layer. That is to be considered as liquid two being dewetted everywhere. Therefore, we expect to retrieve the single layer TFE for the thickness of the lower layer $h_1$ and the upper layer to be constant $h_2\equiv h^*$. 

We have $\lim_{h\to h^*}\varphi_{1}'(h)=0$. Additionally, $\lim_{h\to h^*}\varphi_{1}(h)=1$. With this we calculate  
\begin{subequations}
        \begin{align}
        \begin{split}
            &\lim_{z_2\to z_1+h^*} (p_1-p_2)\\
            &=(\gamma_{s2}-\gamma_{s1}-\gamma_{12})\varphi'(z_1)- (\gamma_{1v}-\gamma_{2v})\nabla^2 z_1
        \end{split} \\
        \begin{split}
            &\lim_{z_2\to z_1+ h^*}(p_2-p_{atm})\\
            &=(\gamma_{12}+ \gamma_{sv}-\gamma_{1v}-\gamma_{2s})\varphi'(z_1) -\gamma_{2v}\nabla^2 z_1.
        \end{split}
        \end{align}
   \end{subequations}
Furthermore, we calculate the limit of the mobility 
\begin{align}
    \lim_{z_2\to z_1+h^*}M=\begin{pmatrix}
        \frac{z_1^3}{3\mu_1} & \frac{z_1^3}{3\mu_1}\\
        0 & 0
    \end{pmatrix}
\end{align}
As such, we see that performing the limit of the two-layer TFE Eq.~\eqref{eq:thin_film_equation_2} that $h_1$ fulfills the single layer TFE Eq.~\eqref{eq:thin_film_equation_1} with $\gamma=\gamma_{13}$, $\gamma_{sl}=\gamma_{01}$, and $\gamma_{sv}=\gamma_{03}$. 

Further, the above calculations for the limit $z_2\to z_1+h^*$ result in $\partial_t h_2=0$.

\subsubsection{$z_1\to h^*$}

The limit $z_1\to h^*$ is the lower liquid layer to be dewetted everywhere. Therefore, we expect the upper layer's thickness $h_2$ to fulfill the single layer TFE Eq.~\eqref{eq:thin_film_equation_1} while $h_1\equiv h^*$. 

We calculate the limit of the pressure jumps to be 
\begin{subequations}
        \begin{align}
        \lim_{z_1\to h^*} (p_1-p_2)=& -(\gamma_{1v}- \gamma_{12}-\gamma_{2v})\varphi'(z_2)\\
        \lim_{z_1\to h^*}(p_2-p_{atm})=&(\gamma_{sv}- \gamma_{2v}-\gamma_{s2})\varphi'(z_2)- \gamma_{2v}\nabla^2z_2.
        \end{align}
\end{subequations}
The limit of the mobility matrix evaluates to
\begin{align}
         \lim_{z_1\to h^*} M = \begin{pmatrix}
           0 & 0 \\
           0 & \frac{z_2^3}{3\mu_2}
       \end{pmatrix}.
   \end{align}
Thus, we see as expected that the two-layer TFE Eq.~\eqref{eq:thin_film_equation_2} becomes in the limit $z_1\to h^*$ the single layer TFE Eq.~\eqref{eq:thin_film_equation_1} for the thickness of the upper layer $h_2$ with surface tensions $\gamma=\gamma_{23}$, $\gamma_{sl}=\gamma_{02}$, and $\gamma_{sv}=\gamma_{03}$, while $\partial_t h_1=0$. 


%

\end{document}